\documentclass[iop]{emulateapj}
\usepackage{natbib,graphicx,amsmath}

\newcommand{\caii}{\ion{Ca}{2} H \& K}

\newcommand{\vsini}{$v \sin i$}
\newcommand{\sval}{\ensuremath{S_{\mbox{\scriptsize HK}}}}

\shorttitle{A \emph{Spitzer} Transmission Spectrum for GJ 436b}
\shortauthors{Knutson et al.}\def\simgr{\,\hbox{\hbox{$ > $}\kern -0.8em \lower 1.0ex\hbox{$\sim$}}\,}
\def\simle{\,\hbox{\hbox{$ < $}\kern -0.8em \lower 1.0ex\hbox{$\sim$}}\,}

\begin{document}

\title{A \emph{Spitzer} Transmission Spectrum for the Exoplanet GJ 436b, Evidence for Stellar Variability, and Constraints on Dayside Flux Variations} 

\author{
Heather A. Knutson\altaffilmark{1,2}, Nikku Madhusudhan\altaffilmark{3}, Nicolas B. Cowan\altaffilmark{4}, Jessie L. Christiansen\altaffilmark{5}, Eric Agol\altaffilmark{6}, Drake Deming\altaffilmark{7}, Jean-Michel D\'esert\altaffilmark{8}, David Charbonneau\altaffilmark{8}, Gregory W. Henry\altaffilmark{9}, Derek Homeier\altaffilmark{10}, Jonathan Langton\altaffilmark{11,12}, Gregory Laughlin\altaffilmark{10}, \& Sara Seager\altaffilmark{13}
}

\altaffiltext{1}{Department of Astronomy, University of California, Berkeley, CA 94720-3411, USA} 
\altaffiltext{2}{Miller Fellow; hknutson@berkeley.edu}
\altaffiltext{3}{Department of Astrophysical Sciences, Princeton University, Princeton, NJ 08544, USA}
\altaffiltext{4}{CIERA, Northwestern University, 2131 Tech Drive, Evanston, IL 60208}
\altaffiltext{5}{NASA Ames Research Center, Moffett Field, CA 94035, USA}
\altaffiltext{6}{Dept. of Astronomy, Box 351580, University of Washington, Seattle, WA 98195, USA}
\altaffiltext{7}{NASA Goddard Space Flight Center, Planetary Systems Laboratory, Code 693, Greenbelt, MD 20771, USA}
\altaffiltext{8}{Harvard-Smithsonian Center for Astrophysics, 60 Garden St., Cambridge, MA 02138, USA}
\altaffiltext{9}{Center of Excellence in Information Systems, Tennessee State University, 3500 John A. Merritt Blvd., Box 9501, Nashville, TN 37209}
\altaffiltext{10}{Institut fŸr Astrophysik, Georg-August-UniversitŠt, Friedrich-Hund-Platz 1, 37077 Gšttingen, Germany}
\altaffiltext{11}{Department of Astronomy and Astrophysics, University of California, Santa Cruz, CA 95064, USA}
\altaffiltext{12}{Department of Physics, Principia College, 1 Maybeck Place, Elsah, IL 62028, USA}
\altaffiltext{13}{Department of Earth, Atmospheric, and Planetary Sciences, Department of Physics, Massachusetts Institute of Technology, Cambridge MA 02139, USA}

\begin{abstract}

In this paper we describe a uniform analysis of eight transits and eleven secondary eclipses of the extrasolar planet GJ 436b obtained in the 3.6, 4.5, and 8.0~\micron~bands using the IRAC instrument on the \emph{Spitzer Space Telescope} between UT 2007 June 29 and UT 2009 Feb 4.  We find that the best-fit transit depths for visits in the same bandpass can vary by as much as 8\% of the total ($4.7\sigma$~significance) from one epoch to the next.  Although we cannot entirely rule out residual detector effects or a time-varying, high-altitude cloud layer in the planet's atmosphere as the cause of these variations, we consider the occultation of active regions on the star in a subset of the transit observations to be the most likely explanation.  We find that for the deepest 3.6~\micron~transit the in-transit data has a higher standard deviation than the out-of-transit data, as would be expected if the planet occulted a star spot.  We also compare all published transit observations for this object and find that transits observed in the infrared typically have smaller timing offsets than those observed in visible light.  In this case the three deepest \emph{Spitzer} transits are all measured within a period of five days, consistent with a single epoch of increased stellar activity.  We reconcile the presence of magnetically active regions with the lack of significant visible or infrared flux variations from the star by proposing that the star's spin axis is tilted with respect to our line of sight, and that the planet's orbit is therefore likely to be misaligned.  In contrast to the results reported by \citet{beaulieu10}, we find no convincing evidence for methane absorption in the planet's transmission spectrum.  If we exclude the transits that we believe to be most affected by stellar activity, we find that we prefer models with enhanced CO and reduced methane, consistent with GJ~436b's dayside composition from \citet{stevenson10}.  It is also possible that all transits are significantly affected by this activity, in which case it may not be feasible to characterize the planet's transmission spectrum using broadband photometry obtained over multiple epochs.  These observations serve to illustrate the challenges associated with transmission spectroscopy of planets orbiting late-type stars; we expect that other systems, such as GJ 1214, may display comparably variable transit depths.  We compare the limb-darkening coefficients predicted by \texttt{PHOENIX} and \texttt{ATLAS} stellar atmosphere models, and discuss the effect that these coefficients have on the measured planet-star radius ratios given GJ~436b's near-grazing transit geometry.  Our measured 8~\micron~secondary eclipse depths are consistent with a constant value, and we place a $1\sigma$ upper limit of 17\% on changes in the planet's dayside flux in this band.  These results are consistent with predictions from general circulation models for this planet, which find that the planet's dayside flux varies by a few percent or less in the 8~\micron~band.  Averaging over the eleven visits gives us an improved estimate of $0.0452\%\pm0.0027\%$ for the secondary eclipse depth; we also examine residuals from the eclipse ingress and egress and place an upper limit on deviations caused by a nonuniform surface brightness for GJ~436b.  We combine timing information from our observations with previously published data to produce a refined orbital ephemeris, and determine that the best-fit transit and eclipse times are consistent with a constant orbital period.  We find that the secondary eclipse occurs at a phase of $0.58672\pm0.00017$, corresponding to $e\cos(\omega)=0.13754\pm0.00027$ where $e$ is the planet's orbital eccentricity and $\omega$ is the longitude of pericenter.  We also present improved estimates for other system parameters, including the orbital inclination, $a/R_{\star}$, and the planet-star radius ratio.

\end{abstract}


\section{Introduction}\label{intro}

Transiting planet systems have proven to be a powerful tool for studying exoplanetary atmospheres.  Observations of transiting systems have been used to detect the signatures of atomic and molecular absorption features at wavelengths ranging from the UV to the infrared \citep[e.g.,][]{charbonneau02,vidal03,swain08,desert08,pont08a,linsky10}, although sometimes the results have proven to be controversial \citep[e.g.,][]{gibson10}.  They have enabled studies of the dayside emission spectra and pressure-temperature profiles of close-in planets \citep[e.g.,][]{charbonneau05,deming05,knutson08,grillmair08}, and they have informed us about their atmospheric circulation \citep[e.g.,][]{knutson07,knutson09a,cowan07,crossfield10}.  Although we currently know of 101 transiting planet systems, our knowledge of these planets (including a majority of the studies cited above) has so far been dominated by studies of the brightest and closest handful of systems, including HD 209458b and HD 189733b.  Planets orbiting small stars offer additional advantages, as they produce proportionally deeper transits and secondary eclipses as a result of their favorable radius ratios and lower stellar effective temperatures.  By this standard, GJ~436 \citep{butler04,maness07,gillon07a,gillon07b,deming07,demory07,torres07} represents an ideal target, as the primary in this system is an early M star with a K band magnitude of 6.1.  

GJ~436b is currently one of the smallest known transiting planets, with a mass only 22 times that of the Earth \citep{torres07}.  Of the planets orbiting stars brighter than 9th magnitude in K band,  only GJ 1214b \citep{charbonneau09}, which also orbits a nearby M dwarf, is smaller.  New discoveries of low-mass transiting planets from space-based surveys such as the \emph{Kepler} and CoRoT missions are unlikely to change this picture significantly, as both include relatively few bright stars.  GJ 436b is also one of the coolest known transiting planets, with a dayside effective temperature of only 800~K \citep{stevenson10}.  Like GJ~1214b, GJ~436b has a high average density indicative of a massive rocky or icy core.  In GJ~436b's case, models indicate that it must also maintain $1-3$ M$_\earth$ of its mass in the form of a H/He atmosphere \citep{adams08,figueira09,rogers10,nettelmann10} in order to match the observed radius.  

By measuring the wavelength-dependent transit depth as GJ~436b passes in front of its host star we can study its atmospheric composition at the day-night terminator, which should be dominated by methane, water, and carbon monoxide \citep{spiegel10,stevenson10,shabram10,madhu10}.  \citet{pont08} observed two transits of GJ~436b with NICMOS grism spectrograph on the \emph{Hubble Space Telescope} (\emph{HST}) and placed an upper limit on the amplitude of the predicted water absorption feature between $1-2$~\micron.  More recently \citet{beaulieu10} reported the detection of strong methane absorption in the 3.6, 4.5, and 8.0~\micron~\emph{Spitzer} bands.  

We can compare these results to observations of the planet's dayside emission spectrum, obtained by measuring the depth of the secondary eclipse when the planet passes behind the star.  \citet{stevenson10} measured secondary eclipse depths for GJ~436b in the 3.6, 4.5, 5.8, 8.0, 16, and 24~\micron~\emph{Spitzer} bands, from which they concluded that the planet's dayside atmosphere contained significantly less methane and more CO than the equilibrium chemistry predictions.  In this work we present an analysis of eight transits and eleven secondary eclipses of GJ~436b observed with \emph{Spitzer}, including an independent analysis of the transit data described in Beaulieu et al., and discuss the corresponding implications for GJ 436b's atmospheric composition.  

Unlike most close-in planets, which typically have circular orbits, GJ~436b has an orbital eccentricity of approximately $0.15$ \citep{maness07,deming07,demory07,madhu09a}.  Atmospheric circulation models for eccentric Jovian planets suggest that they may exhibit significant temperature variations from one orbit to the next \citep{langton08,iro10}, although \citet{lewis10} find little evidence for significant temporal variability in general circulation models for GJ~436b.  The extensive nature of our data set, which includes eleven secondary eclipse observations in the same bandpass obtained between $2007-2009$,  allows us to test the predictions of these models by searching for changes in the planet's 8~\micron~dayside emission on timescales ranging from weeks to years.  

It has also been suggested \citep{ribas08} that GJ~436b's orbital parameters are changing in time, perhaps through perturbations by an unseen second planet in the system.  Such a planet could serve to maintain GJ~436b's nonzero eccentricity despite ongoing orbital circularization, and would not necessarily produce transit timing variations large enough to be detected by earlier, ground-based studies \citep{batygin09}.  Although more recent studies \citep{alonso08,bean08,coughlin08,madhu09a,caceres09,shporer09,ballard10a} have failed to find any evidence for either time-varying orbital parameters or a second transiting object in the system, \emph{Spitzer}'s unparalleled sensitivity and stability allow us to extend the current baseline by nine months with new, high-precision transit observations.

\begin{deluxetable*}{lrrrrcrrrrr}
\tabletypesize{\scriptsize}
\tablecaption{\emph{Spitzer} Observations of GJ 436b \label{obs_table}}
\tablewidth{0pt}
\tablehead{
\colhead{UT Date} & \colhead{Event} & \colhead{\phantom{00} $\lambda$ (\micron)}  & \colhead{Duration (hr)} & \colhead{t$_{int}$ (s)} & \colhead{N$_{exposures}$} & \colhead{Bkd (MJy Sr$^{-1}$)\tablenotemark{a}} & \colhead{Flux (MJy Sr$^{-1}$)\tablenotemark{a,b}} & \colhead{$\sigma_{resid}$\tablenotemark{c}} }
\startdata
UT 2007 Jun 29 & Transit & 8.0\phantom{0} & 3.4\phantom{000} & 0.4 & 28,480 & 530.8\phantom{000} & 9,149.0\phantom{000} & 0.500\% \\
UT 2007 Jun 30 & Eclipse & 8.0\phantom{0} & 5.9\phantom{000} & 0.4 & 49,920 & 544.2\phantom{000} & 9,148.2\phantom{000}   & 0.498\% \\
UT 2008 Jun 11 & Eclipse & 8.0\phantom{0} & 3.4\phantom{000} & 0.4 & 28,800 & 226.0\phantom{000} & 9,156.6\phantom{000} & 0.497\% \\
UT 2008 Jun 13 & Eclipse & 8.0\phantom{0} & 3.4\phantom{000} & 0.4 & 28,800 & 255.0\phantom{000} & 9,161.3\phantom{000} & 0.502\% \\
UT 2008 Jun 16 & Eclipse & 8.0\phantom{0} & 3.4\phantom{000} & 0.4 & 28,800 & 296.5\phantom{000} & 9,154.5\phantom{000} & 0.494\% \\
UT 2008 Jun 19 & Eclipse & 8.0\phantom{0} & 3.4\phantom{000} & 0.4 & 28,800 & 306.4\phantom{000} & 9,151.7\phantom{000} & 0.493\% \\
UT 2008 Jul 12\tablenotemark{d} & Eclipse & 8.0\phantom{0} & 70\phantom{000} & 0.4 & 588,480 & 669.2\phantom{000} & 9,159.9\phantom{000} & 0.506\% \\
UT 2008 Jul 14\tablenotemark{d} & Transit & 8.0\phantom{0} & 70\phantom{000} & 0.4 & 588,480 & 695.6\phantom{000} & 9,160.4\phantom{000} & 0.509\% \\
UT 2008 Jul 15\tablenotemark{d} & Eclipse & 8.0\phantom{0} & 70\phantom{000} & 0.4 & 588,480 & 714.4\phantom{000} & 9,158.1\phantom{000} & 0.515\% \\
UT 2009 Jan 9 & Transit & 3.6\phantom{0} & 4.3\phantom{000} & 0.1 & 117,056 & 37.7\phantom{000} & 36,164.3\phantom{000} & 0.387\% \\
UT 2009 Jan 17 & Transit & 4.5\phantom{0} & 4.3\phantom{000} & 0.1 & 117,056 & 61.6\phantom{000} & 24,382.6\phantom{000} & 0.561\% \\
UT 2009 Jan 25 & Transit & 8.0\phantom{0} & 4.3\phantom{000} & 0.4 & 35,904 & 474.5\phantom{000} & 9,151.9\phantom{000} & 0.502\% \\
UT 2009 Jan 27 & Eclipse & 8.0\phantom{0} & 3.4\phantom{000} & 0.4 & 28,800 & 455.1\phantom{000} & 9,164.5\phantom{000} & 0.495\% \\
UT 2009 Jan 28 & Transit & 3.6\phantom{0} & 4.3\phantom{000} & 0.1 & 117,056 & 82.5\phantom{000} & 36,744.5\phantom{000} & 0.389\% \\
UT 2009 Jan 29 & Eclipse & 8.0\phantom{0} & 3.4\phantom{000} & 0.4 & 28,800 & 411.8\phantom{000} & 9,161.7\phantom{000} & 0.496\% \\
UT 2009 Jan 30 & Transit & 4.5\phantom{0} & 4.3\phantom{000} & 0.1 & 117,056 & 86.3\phantom{000} & 24,177.0\phantom{000} & 0.567\% \\
UT 2009 Feb 1 & Eclipse & 8.0\phantom{0} & 3.4\phantom{000} & 0.4 & 28,800 & 395.9\phantom{000} & 9,163.9\phantom{000} & 0.499\% \\
UT 2009 Feb 2 & Transit & 8.0\phantom{0} & 4.3\phantom{000} & 0.4 & 35,904 & 393.6\phantom{000} & 9,143.8\phantom{000} & 0.501\% \\
UT 2009 Feb 4 & Eclipse & 8.0\phantom{0} & 3.4\phantom{000} & 0.4 & 28,800 & 376.6\phantom{000} & 9,154.5\phantom{000} & 0.499\% \\
\enddata
\tablenotetext{a}{Average sky backgrounds and stellar fluxes estimated for a 5 pixel aperture.}
\tablenotetext{b}{In order to minimize the effects of the detector ramp in the 8.0~\micron~observations we estimate the out-of-transit flux using data after the end of the eclipse event where the ramp is generally smallest; for consistency we use the same region to estimate the fluxes at 3.6 and 4.5~\micron.  We use a 5.0 pixel aperture for the photometry at [3.6,4.5,8.0]~\micron~and apply the appropriate aperture correction of [1.049,1.050,1.068] from Table 4.7 of the IRAC Instrument Handbook to determine the total flux from the star in each observation.}
\tablenotetext{c}{Standard deviation of residuals after dividing out best-fit corrections for instrument effects and transit light curves.}
\tablenotetext{d}{These events were observed as part of a single, continuous phase curve observation with a duration of 70 hours spanning two secondary eclipses and one transit.}
\end{deluxetable*} 

\section{Observations}\label{obs}

We analyze nineteen separate observations of GJ 436, including two 3.6~\micron~transits, two 4.5~\micron~transits, four 8~\micron~transits, and eleven 8~\micron~secondary eclipses, as listed in Table \ref{obs_table}.  All observations were obtained between 2007 and 2009 using the IRAC instrument \citep{faz04} on the \emph{Spitzer Space Telescope} \citep{wern04} in subarray mode.  Some of these data were previously published by other groups, including a transit and secondary eclipse observed on UT 2007 Jun 29/30 \citep{deming07,gillon07a,demory07} and six transits observed in 2009 \citep{beaulieu10}.  Because the two shorter wavelength IRAC channels (3.6 and 4.5~\micron) use InSb detectors and the two longer wavelength channels (5.8 and 8.0~\micron) use Si:As detectors, each of which display different detector  effects, we describe our analysis for each type separately below.  

We calculate the BJD\_UTC values at mid-exposure for each image using the DATE\_OBS keyword in the image headers and the position of \emph{Spitzer}, which is in an earth-trailing orbit, as determined using the JPL Horizons ephemeris.  Each set of 64 images obtained in subarray mode comes as a single FITS file with a time stamp corresponding to the start of the first image; we calculate the time stamps for individual images assuming uniform spacing and using the difference between the AINTBEG and ATIMEEND headers, which record the start and end of the 64-image series.  We then use the routines described in \citet{eastman10} to convert from \emph{Spitzer} JD to BJD\_UTC.  Eastman et al. further advocate a conversion from UTC to TT timing standards, which provide a more consistent treatment of leap seconds.  We note that for the dates spanned by these observations the conversion from BJD\_UTC to BJD\_TT simply requires the addition of [65.184,65.184,66.184]~s for data obtained in [2007,2008,2009], and we leave the dates listed in Table \ref{transit_param} in BJD\_UTC for consistency with other studies.

\subsection{3.6 and 4.5~\micron~Photometry}\label{short_phot}

GJ 436 has a $K$ band magnitude of 6.07, and as a result we elect to use short 0.1~s exposures at 3.6 and 4.5~\micron~in order to ensure that we remain well below saturation.  Subarray images have dimensions of $32\times32$ pixels, making it challenging to estimate the sky background independent of contamination from the wings of the star's point spread function.  We choose to exclude pixels within a radius of 12 pixels of the star's position, as well as the 14th-17th rows, which contain a horizontal diffraction spike that extends close to the edges of the array.  We also exlcude the top (32nd) row of pixels, which have values that are consistently lower than those for the rest of the array.  We then iteratively trim 3$\sigma$~outliers from the remaining subset of approximately six hundred pixels, create a histogram of the remaining values, and then fit a Gaussian to this histogram to determine the sky background for each image.  We find that the background is $0.1-0.2$\% and $0.3-0.4$\% of the total flux in a 5 pixel aperture for the 3.6 and 4.5~\micron~arrays, respectively.

\begin{figure*}
\epsscale{1.1}
\plotone{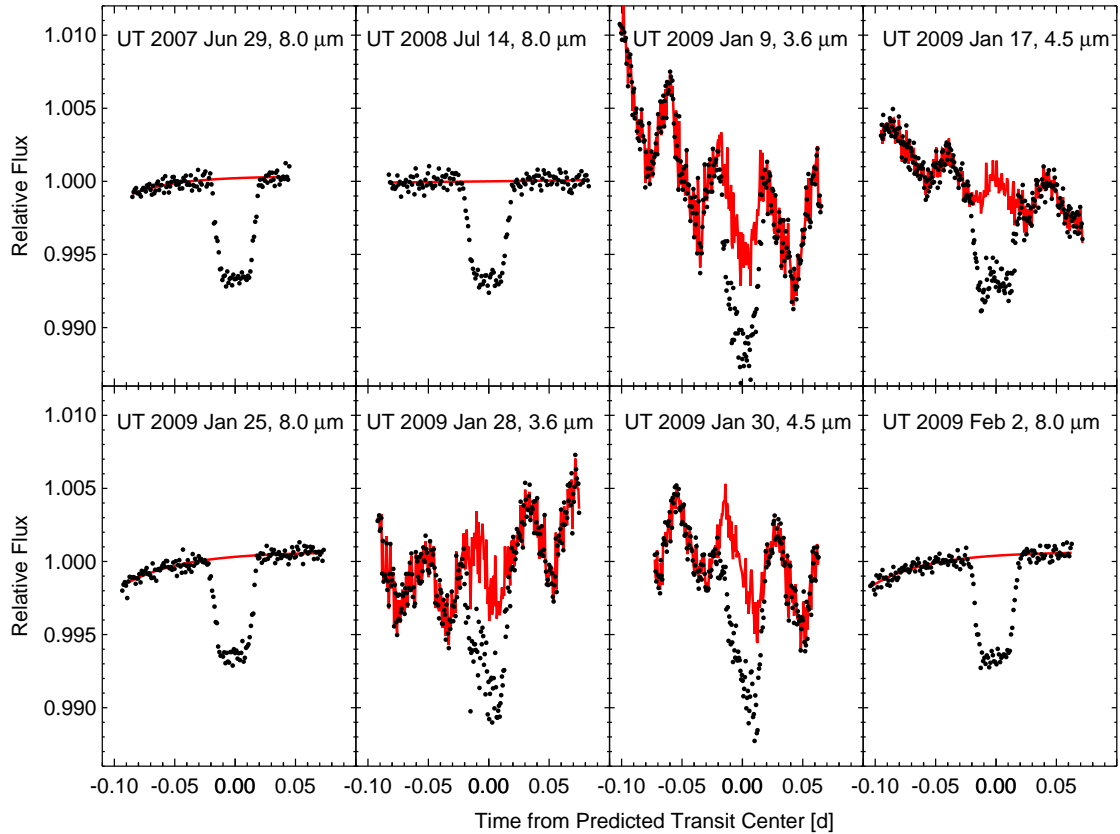}
\caption{Raw photometry for the eight observed transits of GJ 436b, arranged in chronological order and with best-fit detector functions overplotted (solid red lines).  Data has been binned in either 0.9 minute (3.6, 4.5~\micron) or 1.5 minute (8.0~\micron) bins. }
\label{transit_phot_raw}
\end{figure*}

We correct for transient hot pixels by taking a 10-pixel running median of the fluxes at a given pixel position within each set of 64 images and replacing outliers greater than $4\sigma$ with the median value.  We found that using a wider median filter or tighter upper limit for discriminating outliers increased the scatter in the final time series while failing to significantly reduce the number of images that are ultimately discarded.  We find that approximately $0.4-0.8$\% of our images have one or more pixels flagged as outliers using this filter.

Several recent papers have investigated optimal methods for estimating the position of the star on the array for \emph{Spitzer} photometry, with the most extensive discussions appearing in \citet{stevenson10} and \citet{agol10}.  These papers conclude that flux-weighted centroiding \citep[e.g.,][]{knutson08,charbonneau08} and parabola-fitting routines \citep[e.g.,][]{deming06,deming07} tend to produce less than optimal results, while Gaussian fits and least asymmetry methods appear to have fewer systematic biases and a lower overall scatter.  We confirm that for all three wavelengths we obtain better results (defined as a lower scatter in the final trimmed light curve after correcting for detector effects) with Gaussian fits than with flux-weighted centroiding, with a total reduction of $2-7$\% in the standard deviation of the final time series binned in sets of 64 images.  We obtain the best results in both the 3.6 and 4.5~\micron~bands when we first subtract the best-fit background flux from each image, correct bad pixels as described above, and then fit a two-dimensional Gaussian function to a circular region with a radius of 4 pixels centered on the position of the star.  Using smaller or larger fitting regions does not significantly alter the time series but does result in a slightly higher scatter in the normalized light curve.  Although error arrays are available as part of the standard \emph{Spitzer} pipeline, we find that in this case we obtain better results using uniform error weighting for individual pixels.  We use a radially symmetric Gaussian function and run our position estimation routines twice, once where the width is allowed to vary freely in the fits and a second time where we fix the width to the median value over the time series.  Reducing the degrees of freedom by fixing the width produces fits that converge more consistently, with a corresponding improvement in the standard deviation of the normalized time series and fewer large outliers.  \citet{stevenson10} report that they obtain better position estimates when fitting Gaussians to images that have been interpolated to $5\times$ higher resolution, but we find that using interpolated images for our position fits resulted in a slight increase in the scatter in our final light curves.

We perform aperture photometry on our images using the position estimates derived from our Gaussian fits; we expect that aperture photometry will produce the optimal results in light of the low background flux at these wavelengths.  We use apertures with radii ranging between 2.5-7.0 pixels in half pixel steps.  We find that apertures smaller than 3.5 pixels show excess noise, likely connected to position-dependent flux losses, while apertures larger than 5 pixels are more likely to include transient hot pixels and higher background levels, resulting in a higher root-mean-square variance in the final light curve.  We use a 3.5 pixel aperture for our final analysis, but we find consistent results for apertures between $3.5-5.0$ pixels.  We trim outliers from our final time series using a 50 point running median, where we discard outliers greater than $3\sigma$, approximately $2\%$ of the points in a typical light curve.  We find that we trim fewer points when we use flux-weighted centroiding for our position estimates (typically $0.6\%$), but the uncertainties in our best-fit transit parameters are still larger than with the Gaussian fits due to the increased scatter in the final trimmed time series.  We also trim the first 15 minutes in all observations except for the  4.5~\micron~transit on UT 2009 Jan 30, where we trim the first hour of data.  Images taken at the start of a new observation tend to have larger pointing offsets, most likely due to the settling time of the telescope at the new pointing; we find that discarding these early data improves the quality of the fit to the subsequent points.   For all visits other than the transit on UT 2009 Jan 30, we find that we achieve consistent results when we trim either the first 15, 30, or 60 minutes of data, and we therefore choose to trim the minimal 15-minute interval.  For the UT 2009 Jan 30 observation we find that the data display an additional time dependence that is not well-described by the standard linear function of time in Eq. \ref{eq1}, but is instead better-described by a linear function of ln($dt$).  This may be due to the fact that the star falls near the edge of a pixel in these observations, which could introduce additional time-dependent effects.  Rather than changing the functional form used to fit these data, we instead opt to trim the first hour of observations, which removes the most steeply-varying part of the time series and leaves a trend that is well-described by the same linear function of time used in the other transit fits.  We find that we obtain the same transit parameters for this visit when we either trim the first 15 minutes of data and fit with a linear function of ln($dt$) or trim the first hour of data and fit with a linear function of time, so this choice does not affect our final conclusions.

\begin{figure*}
\epsscale{1.1}
\plotone{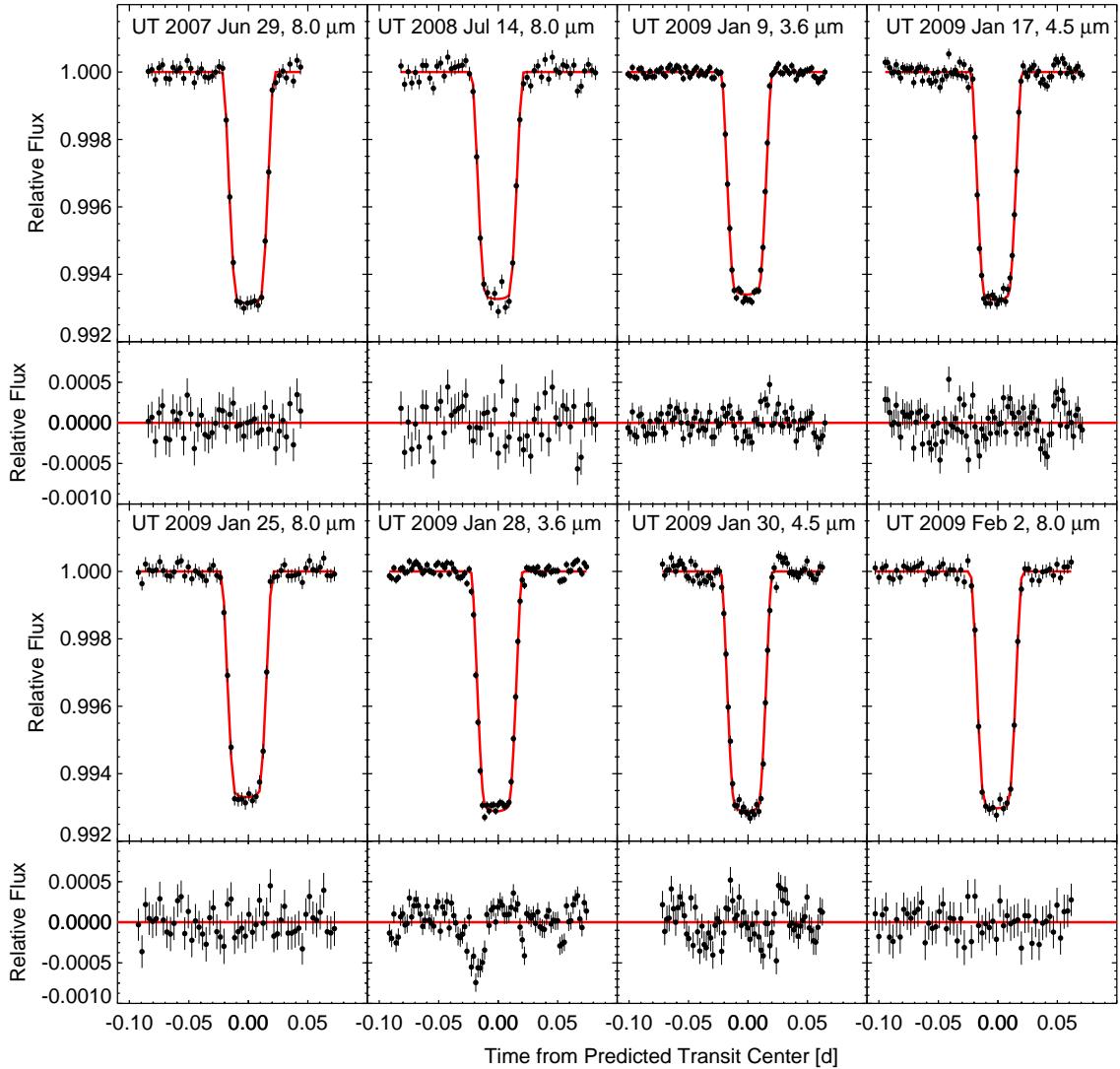}
\caption{Photometry for the eight observed transits of GJ 436b after the best-fit corrections for instrument effects are removed, arranged in chronological order.  Data has been binned in either 2.7 minute (3.6, 4.5~\micron) or 4.3 minute (8.0~\micron) bins. Best-fit transit curves are overplotted in red, and the residuals from each fit are shown in the lower panel.  In this plot we have assumed a constant ephemeris for the planet rather than using the best-fit transit times.  Note that although the out-of-transit residuals for the second 3.6~\micron~observation on UT 2009 Jan 28 appear to be relatively Gaussian, there are additional variations during the transit that are not well-accounted for by the best-fit transit light curve.  These variations are likely due to occultations of spots or faculae by the planet.  The residuals for the 4.5~\micron~transit observed on UT 2009 Jan 30 display excess correlated noise both in and out of transit, most likely due to an imperfect correction for the sharp flux variations caused by the star's location at the edge of a pixel.}
\label{transit_phot_norm}
\end{figure*}

Fluxes measured at these two wavelengths show a strong correlation with the changing position of the star on the array, at a level comparable to the depth of the secondary eclipse.  This effect is due to a well-documented intra-pixel sensitivity variation \citep[e.g.,][]{reach05,charbonneau05,charbonneau08,morales06,knutson08}, in which the sensitivity of an individual pixel differs by several percent between the center and the edge.  The 3.6~\micron~array typically exhibits larger sensitivity variations than the 4.5~\micron~array, as demonstrated by the UT 2009 Jan 9 and 17 transits.  The UT 2009 Jan 30 transit falls very near the edge of a pixel in the 4.5~\micron~subarray, and thus displays a sensitivity variation comparable to that of the more centrally-located 3.6~\micron~transit on UT 2009 Jan 28.  We correct for these sensitivity variations by fitting a quadratic function of $x$ and $y$ position simultaneous with the transit light curve:

\begin{figure*}
\epsscale{1.1}
\plotone{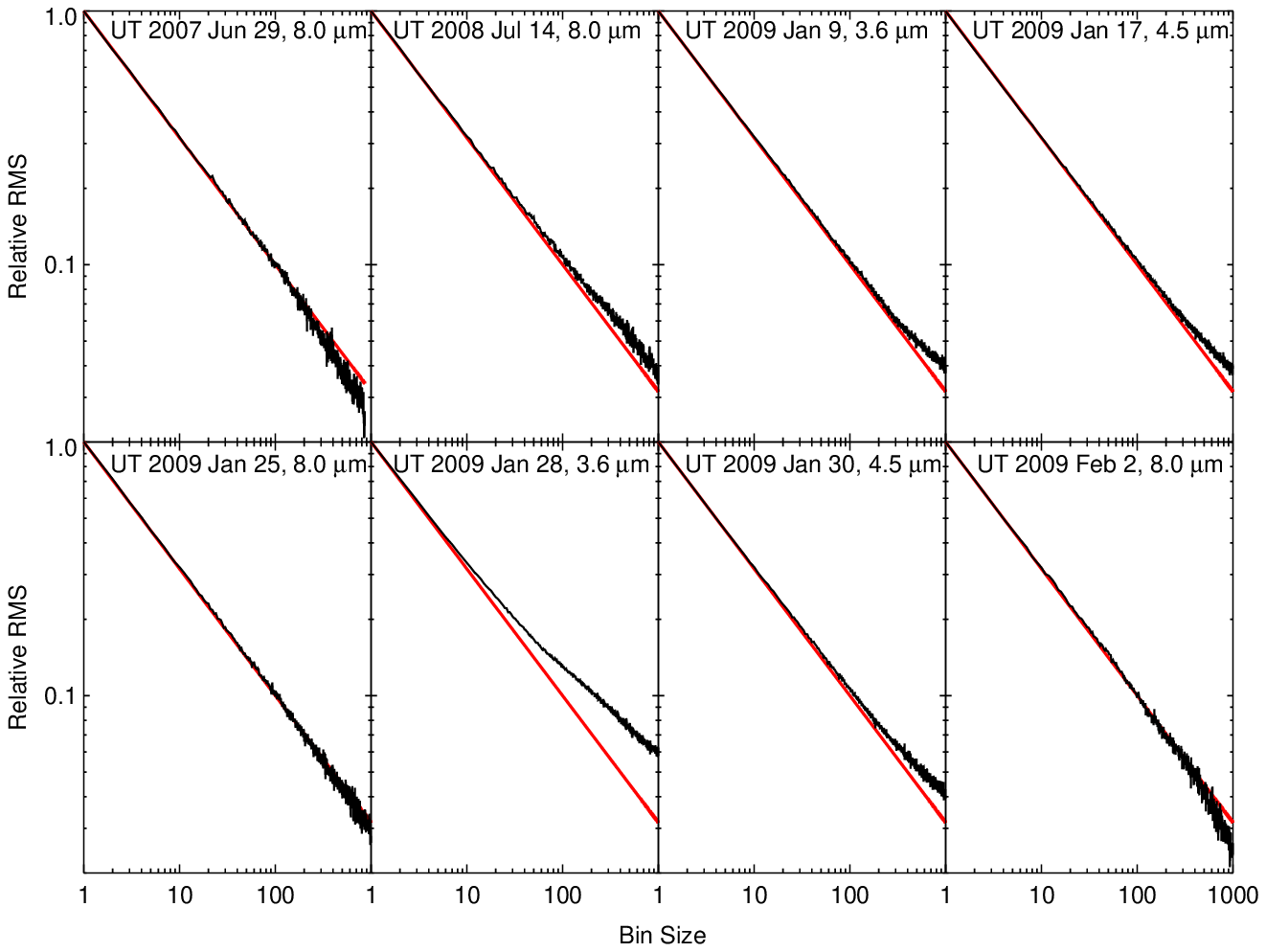}
\caption{Standard deviation of residuals vs. bin size for the eight transits observed with \emph{Spitzer}, arranged in chronological order.  Observations were taken at 8.0, 8.0, 3.6, 4.5~\micron~(top row, left to right), 8.0, 3.6, 4.5, and 8.0~\micron~(bottom row, left to right), respectively. The red curve shows the predicted root-n scaling expected for Gaussian noise.}
\label{transit_rootn}
\end{figure*}

\begin{deluxetable*}{lrrrrcrrrrr}
\tabletypesize{\scriptsize}
\tablecaption{Individual Best-Fit Transit Parameters \label{transit_param}}
\tablewidth{0pt}
\tablehead{
\colhead{UT Date} & \colhead{$\lambda$~(\micron)} & \colhead{$R_p/R_{\star}$} & \colhead{\phantom{aaaa}Depth} & \colhead{\phantom{aaaaa}Transit Center (BJD)}  & \colhead{O-C (s)\tablenotemark{a}} }
\startdata
UT 2007 Jun 29 & 8.0 & $0.08322\pm 0.00052$ & $0.6926\%\pm0.0087\%$ &  $ 2454280.78193\pm0.00012$ & $12.5\pm10.2$ \\
UT 2008 Jul 14 & 8.0 & $0.08247\pm 0.00061$ & $0.6801\%\pm0.0101\%$ & $2454661.50314\pm0.00017$ & $6.2\pm14.4$ \\
UT 2009 Jan 9 & 3.6 & $0.08182\pm 0.00037$ & $0.6694\%\pm0.0061\%$ & $2454841.28821\pm0.00008$ & $6.9\pm6.5$ \\
UT 2009 Jan 17 & 4.5 & $0.08286\pm0.00047$ & $0.6865\%\pm0.0078\%$ & $2454849.21985\pm0.00012$ & $2.1\pm10.5$ \\
UT 2009 Jan 25 & 8.0 & $0.08224\pm0.00051$ & $0.6763\%\pm0.0084\%$ & $2454857.15155\pm0.00012$ & $3.2\pm10.1$ \\
UT 2009 Jan 28 & 3.6 & $0.08495\pm0.00056$ & $0.7216\%\pm0.0095\%$ & $2454859.79504\pm0.00012$ & $-31.9\pm10.3$ \\
UT 2009 Jan 30 & 4.5 & $0.08502\pm0.00057$ & $0.7227\%\pm0.0097\%$ & $2454862.43970\pm0.00011$ & $33.7\pm9.6$ \\
UT 2009 Feb 2 & 8.0 & $0.08424\pm0.00049$ & $0.7096\%\pm0.0083\%$ & $2454865.08345\pm0.00012$ & $20.8\pm10.6$ \\
\enddata
\tablenotetext{a}{Observed minus calculated transit times.  Predictions use the best-fit ephemeris of $T_c=2454865.083208\pm0.000042$ BJD and $P=2.6438979\pm0.0000003$ days from Table \ref{global_param}.}
\end{deluxetable*}

\begin{align}\label{eq1}
f & = & f_0*(a_1+a_2(x-x_0)+a_3(x-x_0)^2 \nonumber \\
& & +a_4(y-y_0)+a_5(y-y_0)^2+a_6t)
\end{align}
where $f_0$ is the original flux from the star, $f$ is the measured flux, $x$ and $y$ denote the location of the star on the array, $x_0$ and $y_0$ are the median $x$ and $y$ positions, $t$ is the time from the predicted eclipse center, and $a_1-a_6$ are free parameters in the fit.  In both bandpasses we find that quadratic terms in both $x$ and $y$ are necessary to achieve a satisfactory fit to the observed variations, although the $\chi^2$ value for the fits is not improved by the addition of an $xy$ term, or higher-order terms in $x$ and $y$.  We find that the fits are also improved by the addition of a linear term in time, consistent with previous observations at these wavelengths \citep[e.g.,][]{knutson09,todorov10,fressin10,odonovan10,deming11}.

\begin{figure*}
\epsscale{1.1}
\plotone{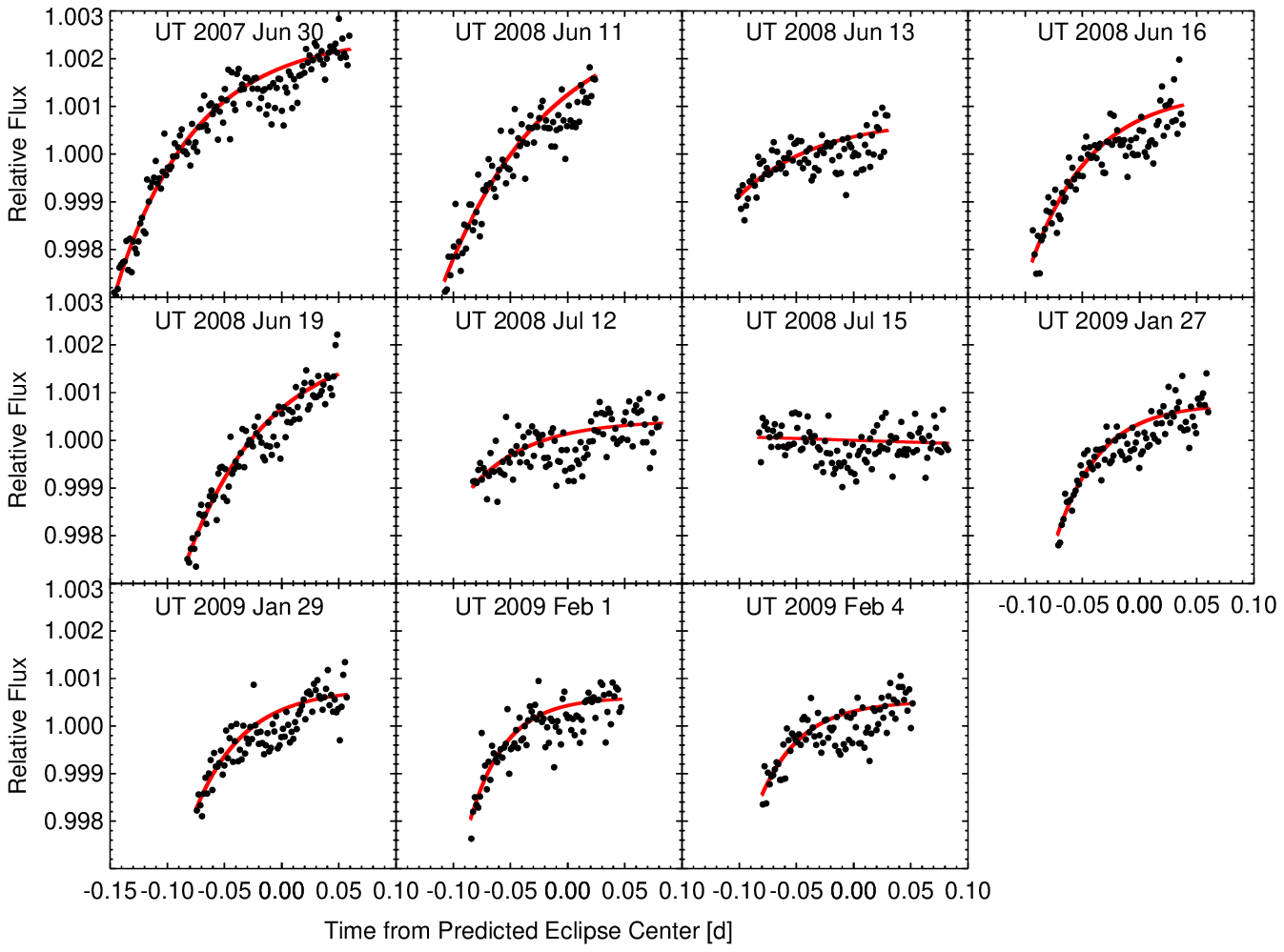}
\caption{Raw photometry for eleven 8~\micron~secondary eclipses of GJ 436b, arranged in chronological order.  Data has been binned in 2.2 minute bins, and the best-fit corrections for detector effects in each visit are overplotted in red.}
\label{eclipse_phot_raw}
\end{figure*}

\subsection{8.0~\micron~Photometry}\label{long_phot} 

We follow the same methods described in \S\ref{short_phot} to estimate the sky background in the 8.0~\micron~images, except in this case we include pixels at distances of more than ten pixels from the position of the star in our estimate instead of the previous twelve-pixel radius.  The background in these images ranges between $2.6-7.7$\% of the total flux in a five pixel aperture, and we find that including pixels between $10-12$ pixels from the star's position improves the accuracy of our background estimates without adding significant contamination from the star's point spread function.  In \citet{agol10} we find that using a slightly larger 4.5~pixel aperture instead of 3.5 pixels minimizes correlated noise in 8~\micron~\emph{Spitzer} observations (albeit at the cost of slightly higher Gaussian noise), and we therefore elect to use a 4.5 pixel aperture for our 8~\micron~data.  Our choice of aperture has a negligible effect on the best-fit eclipse depths and times, as we find consistent results for apertures between $3.5-5.0$ pixels. 

\begin{figure*}
\epsscale{1.1}
\plotone{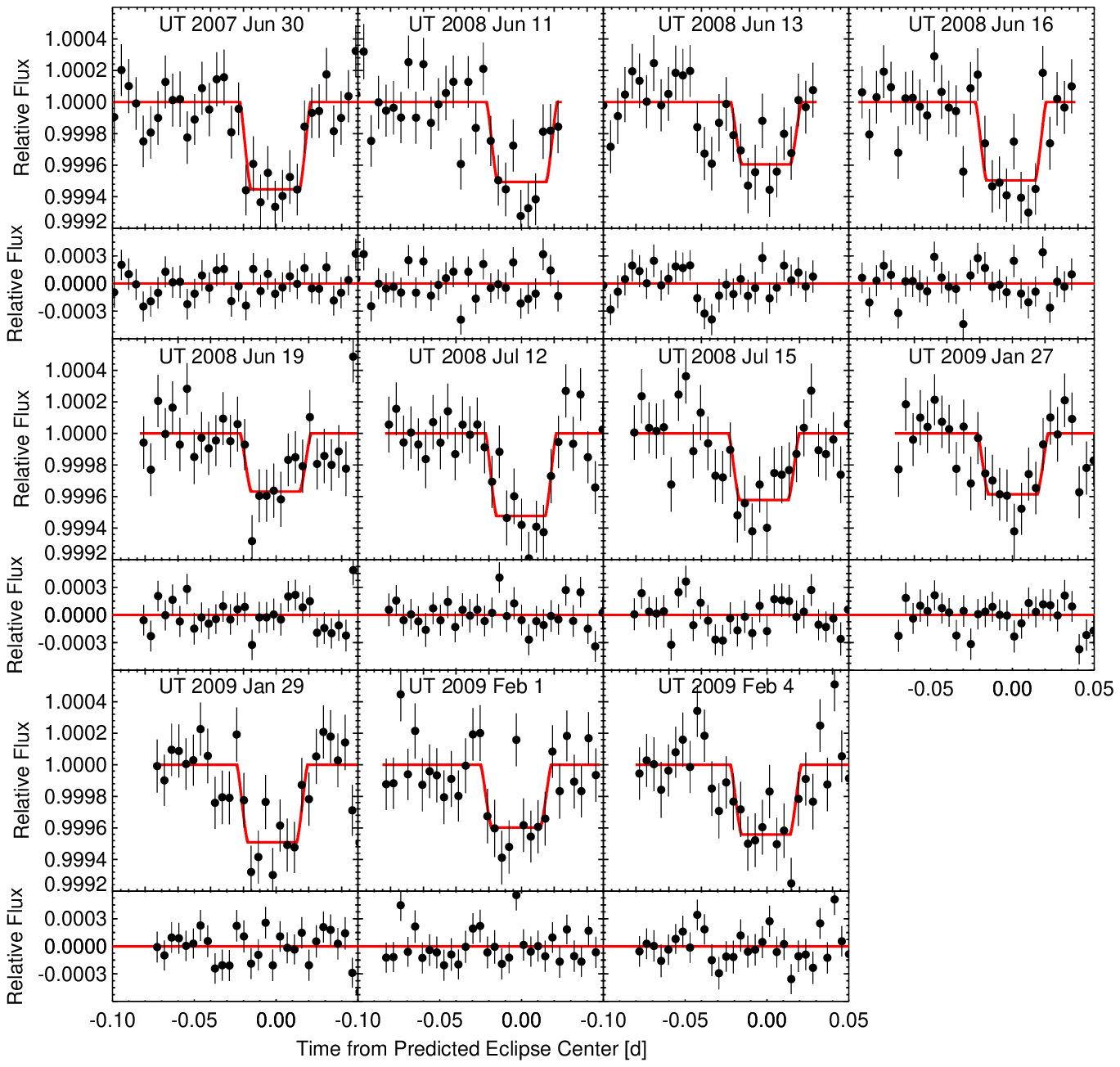}
\caption{Photometry for eleven 8~\micron~secondary eclipses of GJ 436b, arranged in chronological order.  Data has been binned in 6.4 minute bins, and the best-fit eclipse curve for each visit is overplotted in red.  The residuals for each visit are shown in the panels below the eclipses.}
\label{eclipse_phot_norm}
\end{figure*}

\begin{figure*}
\epsscale{1.1}
\plotone{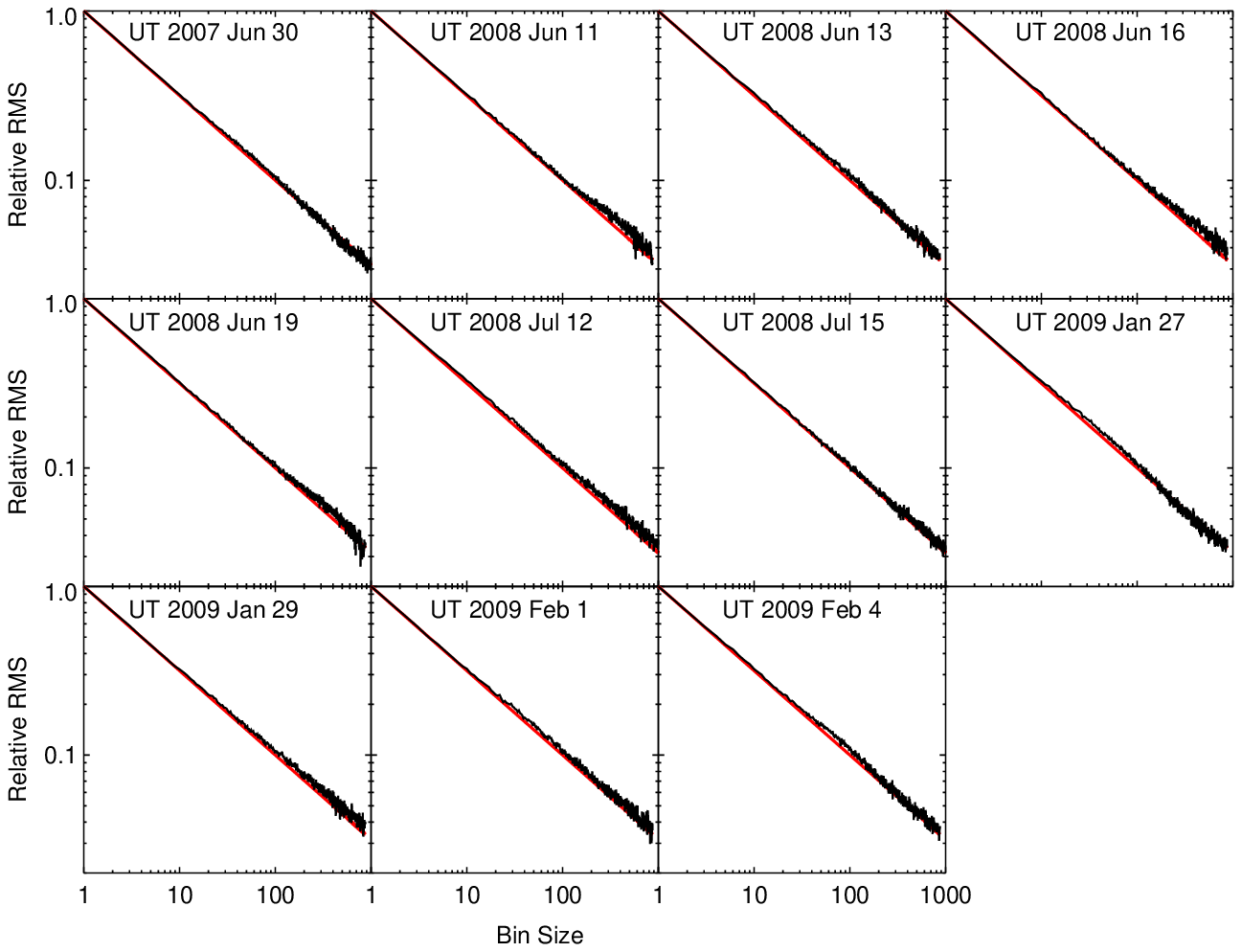}
\caption{Standard deviation of residuals vs. bin size for the eleven secondary eclipses observed with \emph{Spitzer}, arranged in chronological order.   The red curve shows the predicted root-n scaling expected for Gaussian noise, and the lack of any excess noise for large bin sizes suggests that these light curves should be well-described by the standard MCMC error analysis.}
\label{eclipse_rootn}
\end{figure*}

\emph{Spitzer} fluxes for stars observed using the IRAC 8.0~\micron~array, the IRS 16~\micron~array, and the MIPS 24~\micron~array do not appear to have a significant position dependence, but do display a ramp-like behavior where higher-illumination pixels converge to a constant value within the first hour of observations while lower-illumination pixels show a continually increasing linear trend on the time scales of interest here.  This effect is believed to be due to charge-trapping in the array, and is discussed in detail in \citet{knutson07,knutson09c} and \citet{agol10}, among others.  We mitigate this effect in our data by staring either at a bright star (HD 107158 in the case of the 8~\micron~secondary eclipse observations between UT 2008 Jun 11 and Jun 19) or an HII region with bright diffuse emission at 8~\micron~(LBN 543 for the 8~\micron~transit observations, and G111.612+0.374 for the 8~\micron~secondary eclipse observations between UT 2009 Jan 27 and Feb 4) for approximately 30 minutes prior to the start of our observations.  The 2007 observations were obtained prior to the development of this preflash technique, but as discussed in \citet{deming07} the transit observation happened to follow an observation of another bright object and thus was effectively preflashed in the same manner as the 2008 and 2009 data.  The secondary eclipse observed in 2007 was not preflashed, and thus displays a much steeper ramp than the other observations.  We examine the distribution of ramp slopes in Fig. \ref{eclipse_phot_raw} and find no correlation between the relative offsets in the positions of GJ 436 and the preflash star and the slope of the subsequent ramp; the preflash star is offset by [0.4, 0.2, 0.3, 0.1] pixels in the UT 2008 Jun 11, 13, 16, and 19 observations, respectively, but the shallowest ramp occurs in the Jun 13 observation while the smallest offset occurs in the Jun 19 observation.   We speculate that the UT 2008 Jun 13 observation may have been effectively preflashed by the preceding science observations in the same way as the UT 2007 Jun 29 transit observation.  We find that all forms of preflash reduce the slope of the subsequent ramp as compared to the non-preflashed secondary eclipse on UT 2007 June 30, but the HII regions consistently produce a larger reduction in the ramp slope than preflashes using a bright star.  

We can describe the ramps in our 8~\micron~science data with the following functional form:

\begin{align}\label{eq2}
f=c_1\left(1-c_2\exp{}(-\delta t/c_3)-c_4\exp(-\delta t/c_5)\right)
\end{align}
where $f$ is the measured flux, $\delta t$ is the elapsed time from the start of the observations, and $c_1-c_5$ are free parameters in the fit.  Previous studies have elected to use either a single exponential \citep[e.g.,][]{harrington07}, a linear + log function of $\delta t$ \citep[e.g.,][]{deming07}, or a quadratic function in $\log(\delta t)$ \citep[e.g.,][]{charbonneau08,knutson08,desert09}.  However, in \citet{agol10} we find that the functional forms involving $\log(\delta t)$ produce eclipse depths that are correlated with the slope of the observed ramp function, while the single exponential does not provide a good fit to data with a steep ramp.  We conclude that a double exponential function has enough degrees of freedom to fit a range of ramp profiles, while still avoiding correlations between the measured eclipse depths and the slope of the detector ramp.  Although we require a double exponential function in order to fit the steeper, non-preflashed 2007 secondary eclipse observation in this study,  we obtain comparable results with a single exponential term for our preflashed 8~\micron~data.  We therefore elect to use this simpler single exponential in our subsequent analysis for all 8~\micron~visits except the 2007 secondary eclipse.

For our fits to phase curve data obtained on UT 2008 July 12-15, we select a four hour subset of data centered on the position of the transit or eclipse and use that in our fits.  The first eclipse takes place at the start of the observations, which exhibit a residual ramp, and we therefore fit this light curve with the same single exponential as our other data.  We use a linear function of time to fit the out-of-eclipse trends in the transit, which occurs in the middle of the observations, as well as the secondary eclipse towards the end of the observations.  We find that the scatter in the central region of the time series near the transit, when the star is closest to the edge of the pixel, is higher than for either secondary eclipse or for the other 8~\micron~transit observations.  \citep{stevenson10} found that the fluxes measured with the 5.8~\micron~\emph{Spitzer} array sometimes display a weak dependence on the position of the star, which may be due to either flat-fielding errors or intrapixel sensitivity variations similar to those observed in the 3.6 and 4.5~\micron~arrays, although no such effect has been definitively detected in the 8~\micron~array to date.  We test for the presence of position-dependent flux variations in our data by adding linear functions of $x$ and $y$ position to each of our 8~\micron~transit fits, and find that the $\chi^2$ value of the resulting fits is effectively unchanged in all cases except for the UT 2008 July 14 transit, where it decreased from 37,186.6 to 37,177.7 for 33,636 points and six degrees of freedom.  Using the Bayesian Information Criterion described in \citet{stevenson10}, we conclude that this reduction in $\chi^2$ is not significant, and we exclude these position-depedent terms in our subsequent analysis of the 8~\micron~data.

\begin{deluxetable*}{lrrrrcrrrrr}
\tabletypesize{\scriptsize}
\tablecaption{Individual Best-Fit 8~\micron~Secondary Eclipse Parameters \label{eclipse_param}}
\tablewidth{0pt}
\tablehead{
\colhead{UT Date} & \colhead{Depth (\%)} & \colhead{Eclipse Center (BJD)}  & \colhead{O-C (min)\tablenotemark{a}} }
\startdata
UT 2007 Jun 30 & $0.0553\pm0.0083$ &  $2454282.3329\pm0.0016$ & $-0.2\pm2.3$ \\
UT 2008 Jun 11 & $0.0506\pm0.0110$ &  $ 2454628.6850\pm0.0017$ & $2.0\pm2.4$ \\
UT 2008 Jun 13 & $0.0395\pm0.0097$ & $2454631.3281\pm0.0021$ & $0.9\pm3.0$ \\
UT 2008 Jun 16 & $0.0497\pm0.0087$ & $2454633.9716\pm0.0013$ & $0.3\pm1.9$ \\
UT 2008 Jun 19 & $0.0368\pm0.0089$ & $2454636.6162\pm0.0021$ & $1.2\pm3.0$ \\
UT 2008 Jul 12 & $0.0523\pm0.0090$ &  $2454660.4112\pm0.0019$ & $1.2\pm2.8$ \\
UT 2008 Jul 15 & $0.0422\pm0.0078$ & $2454663.0537\pm0.0040$ & $-0.9\pm5.8$ \\
UT 2009 Jan 27 & $0.0386\pm0.0087$ & $2454858.7047\pm0.0026$ & $2.8\pm3.8$ \\
UT 2009 Jan 29 & $0.0491\pm0.0088$ & $2454861.3460\pm0.0015$ & $-1.0\pm2.2$ \\
UT 2009 Feb 1 & $0.0398\pm0.0086$ & $2454863.9889\pm0.0017$ & $-2.4\pm2.4$ \\
UT 2009 Feb 4 & $0.0441\pm0.0087$ & $2454866.6355\pm0.0023$ & $1.4\pm3.3$ \\
\enddata
\tablenotetext{a}{Observed minus calculated transit times.  Predictions use the best-fit ephemeris of $T_c=2454865.083208\pm0.000042$ BJD and $P=2.6438979\pm0.0000003$ days, and an orbital phase of $0.58685\pm0.00017$ from Table \ref{global_param}.}
\end{deluxetable*}

\subsection{Transit and Eclipse Fits}\label{transits}  

We carry out simultaneous fits to determine the best-fit transit functions and detector corrections using a non-linear Levenberg-Marquardt minimization routine \citep{markwardt09}.  We calculate our eclipse curve using the equations from \citet{mand02} assuming a longitude of pericenter equal to $334\degr~\pm10\degr$~(update based on complete set of published and unpublished radial velocity data, A. Howard, personal communication, 2010).  The orbital eccentricity determined from the updated radial velocity data is $0.145\pm0.017$, but we choose to set the orbital eccentricity equal to 0.152 in our fits, which we calculate using the above longitude of pericenter and the published value of $e*cos(\omega)=0.1368\pm0.0004$ from \citet{stevenson10}.  We find that the uncertainty in the calculated eccentricity is dominated by the uncertainty in $\omega$, but this has a minimal impact on our transit fits.  Our best-fit parameters change by less than $1\sigma$ for eccentricity values between 0.142 and 0.169, corresponding to $\pm10\degr$ in $\omega$, where our best-fit inclination is most sensitive to the assumed eccentricity ($0.9\sigma$ change), $a/R_{\star}$ is somewhat sensitive ($0.5\sigma$ change), and the best-fit radius ratios and transit times for individual fits are minimally sensitive ($<0.3\sigma$ change).  Our nominal values for the eccentricity and longitude of pericenter result in a transit length of 60.9 minutes, 0.5 minutes longer than the zero eccentricity case.  Using the same parameters for the secondary eclipse, which occurs shortly before periastron passage, produces a length of 62.6 minutes.

\begin{deluxetable}{lrrrrcrrrrr}
\tabletypesize{\scriptsize}
\tablecaption{Four-Parameter Nonlinear Limb-Darkening Coefficients\tablenotemark{a} \label{limb_dk}}
\tablewidth{0pt}
\tablehead{
\colhead{Model} & \colhead{Band~(\micron)} & \colhead{$c_1$} & \colhead{$c_2$} & \colhead{$c_3$} & \colhead{$c_4$} }
\startdata
\texttt{ATLAS} & 3.6\phantom{band} & 1.122 & -1.852 & 1.675 & -0.582 \\
\texttt{ATLAS} & 4.5\phantom{band} & 0.749 & -0.917 & 0.718 & -0.230 \\
\texttt{ATLAS} & 5.8\phantom{band} & 0.815 & -1.147 & 0.947 & -0.310 \\
\texttt{ATLAS} & 8.0\phantom{band} & 0.770 & -1.141 & 0.942 & -0.304 \\
\texttt{PHOENIX} & 3.6\phantom{band} & 1.284 & -1.751 & 1.433 & -0.470 \\
\texttt{PHOENIX} & 4.5\phantom{band} & 1.203 & -1.796 & 1.512 & -0.500 \\
\texttt{PHOENIX} & 5.8\phantom{band} & 0.918 & -1.264 & 1.064 & -0.358 \\
\texttt{PHOENIX} & 8.0\phantom{band} & 0.619 & -0.762 & 0.645 & -0.220 \\
\enddata
\tablenotetext{a}{Both models assume $T_\mathrm{eff}=3500$~K and [Fe/H]=0.  The \texttt{ATLAS} model uses $\log(g)=5.0$ and the \texttt{PHOENIX} model uses $\log(g)=4.76$ for better consistency with the radius and luminosity in \citet{torres07}, but emipirical tests show the assumed surface gravity has a negligible effect on the resulting limb-darkening profiles.  For a definition of this limb-darkening law, see \citet{claret00}.}
\end{deluxetable}

We fit the eight transits simultaneously and assume that the inclination and the ratio of the orbital semi-major axis to the stellar radius $a/R_{\star}$ are the same for all transits, but allow the planet-star radius ratio $R_p/R_{\star}$ and transit times to vary individually.  Figure \ref{transit_phot_raw} shows the final binned data from these fits with the best fit normalizations for the detector effects and transit light curves in each channel overplotted, and Figure \ref{transit_phot_norm} shows the binned data once these trends are removed, with best-fit transit curves overplotted.  Best-fit parameters are given in Tables \ref{transit_param} and \ref{global_param}.

\subsubsection{A Comparison of \texttt{ATLAS} and \texttt{PHOENIX} Limb-Darkening Models}

We derive limb-darkening coefficients for the star using a Kurucz \texttt{ATLAS} stellar atmosphere model with $T_\mathrm{eff}=3500$~K, $\log(g)=5.0$, and [Fe/H]$=0$ \citep{kurucz79,kurucz94,kurucz05}, where we take the flux-weighted average of the intensity profile in each IRAC band and then fit this profile with four-parameter nonlinear limb-darkening coefficients \citep{claret00}.  We also derive limb-darkening coefficients for a \texttt{PHOENIX} atmosphere model \citep{hauschildt99} with the same parameters, and list both sets of coefficients in Table \ref{limb_dk}.  We trim the maximum stellar radius in the \texttt{PHOENIX} models, which is set to an optical depth of $10^{-9}$, to match the level of the $\tau=1$ surface in each \emph{Spitzer} band.  We estimate the location of this surface by determining when the intensity relative to that at the center of the star first drops below $e^{-1}$, and find that the new stellar radius is $0.09-0.10\%$ smaller than the old $\tau=10^{-9}$ value.  We find that we can achieve satisfactory four-parameter nonlinear fits to the \texttt{PHOENIX} intensity profiles only when we exclude points where $\mu<0.025$, whereas the \texttt{ATLAS} models are well-described by fits including this region.  

\begin{figure}
\epsscale{1.2}
\plotone{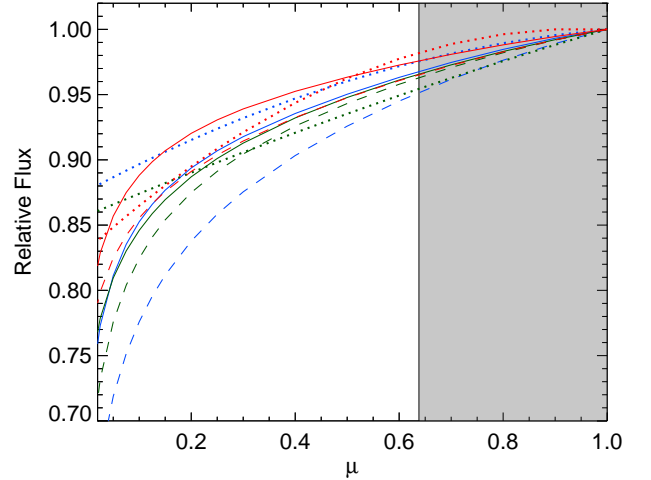}
\caption{A comparison of the model limb-darkening as a function of $\mu=cos(\theta)$, where $\theta$ ranges from $0\degr$ at the center of the star to $90\degr$ at the edge.  We show limb-darkening from four-parameter nonlinear limb-darkening coefficients obtained using \texttt{ATLAS} (solid lines) or \texttt{PHOENIX} (dashed lines) stellar atmosphere models, as well as the best-fit quadratic limb-darkening obtained by a fit to our data (dotted lines).  The color indicates the bandpass, including 3.6~\micron~(blue), 4.5~\micron~(green), and 8.0~\micron~(red) \emph{Spitzer} bands.  The grey shaded region indicates values of $\mu$ for which we have no direct observational constraints, as the planet does not cross this part of the star during its transit.}
\label{limbdk}
\end{figure}

As illustrated in Fig. \ref{limbdk}, the \texttt{PHOENIX} model predicts stronger limb darkening in all bands as compared to the \texttt{ATLAS} model, with the largest differences in the 3.6~\micron~band.  When we compare our best-fit transit parameters using either the \texttt{ATLAS} or \texttt{PHOENIX} limb-darkening coefficients, we find that the best-fit planet-star radius ratios are $0.8-1.2\sigma$ ($0.5-0.6\%$) deeper in the 3.6~\micron~band, $0.06-0.07\sigma$ ($0.04-0.05\%$) smaller in the 4.5~\micron~band, and $0.3-0.4\sigma$ ($0.2-0.3\%$) larger in the 8.0~\micron~band for the \texttt{PHOENIX} models.  The best-fit values for the inclination and $a/R_{\star}$ increase by $1.0\sigma$ ($0.6\%$) and $0.9\sigma$ ($0.04\%$), respectively, for the \texttt{PHOENIX} model fits; this is a product of the stronger limb-darkening profile, as GJ~436b's relatively high impact parameter creates a partial degeneracy between the limb-darkening profile and the other transit parameters.

We examine the relative importance of the assumed stellar parameters by comparing two \texttt{PHOENIX} models with effective temperatures of 3400~K and 3600~K.  We find that for this 200~K range in effective temperature, the best-fit planet-star radius ratios change by $0.11-0.16\sigma$ at 3.6~\micron, $0.07-0.09\sigma$ at 4.5~\micron, and $0.10-0.12\sigma$ at 8.0~\micron.  The changes in the best-fit values for the inclination and $a/R_{\star}$ were similarly small, $0.04\sigma$ and $0.4\sigma$, respectively.  We therefore conclude that changes in the stellar effective temperature of less than 200~K are negligible for the purposes of our transit fits.  We also compute \texttt{PHOENIX} model intensity profiles for $0.0<$[Fe/H]$<+0.3$, but we find that the differences between models are much smaller than for our 200~K change in the effective temperature.  

As there are currently few observational constraints on limb-darkening profiles for main-sequence stars \citep[e.g.,][]{claret08,claret09}, and even fewer constraints for M~stars in the mid-infrared, we also consider simultaneous transit fits in which we allow quadratic limb-darkening coefficients in each band to vary as free parameters.  As a result of the planet's high impact parameter, our observations do not directly constrain the limb-darkened intensity for values of $\theta \le 50\degr$, corresponding to $\mu \ge 0.64$, as the planet does not cross this region on the star.  However, we can infer the limb-darkening profile in this region if we assume a simple quadratic limb-darkening law.  We require the intensity profile computed from these coefficients to be always less than or equal to one (i.e., no limb brightening), and we require the relative intensity at the edge of the star to be greater than or equal to the equivalent $K$ band limb-darkening from \citet{claret00}.  The dotted lines in Fig. \ref{limbdk} show the resulting best-fit limb-darkening profiles in each band; these profiles show less contrast than either model, but the \texttt{ATLAS} models appear to provide the closest match.  

This agreement is reflected in the $\chi^2$ values for the simultaneous transit fits; the total $\chi^2$ for the best-fit quadratic coefficients is 536,729.25, for the 3500~K \texttt{ATLAS} limb-darkening coefficients it is 536,733.98, and for the [3400, 3500, 3600]~K \texttt{PHOENIX} models it is [536,740.69, 536,739.75, 536,738.81], for 536,798 points and either 53 (with fixed limb-darkening) or 59 (with freely varying quadratic limb-darkening coefficients) free parameters.  We use the \texttt{ATLAS} limb-darkening coefficients in our subsequent analysis, as they produce a marginally better agreement with the best fit profiles than the \texttt{PHOENIX} models.  Although the $\chi^2$ value for the best-fit quadratic limb-darkening coefficients is formally smaller than that of either model, this fit also contains six additional degrees of freedom, making the difference negligible.

As an additional test, we also repeat our fits with the limb-darkening coefficients fixed to zero in all bands.  This produces planet-star radius ratios that are $1.6-2.4\sigma$ ($1.1\%$) smaller in the 3.6~\micron~band, $1.7-2.0\sigma$ ($1.1\%$) smaller in the 4.5~\micron~band, and $0.9-1.1\sigma$ ($0.6-0.7\%$) smaller in the 8.0~\micron~band.  The best-fit inclination and $a/R_{\star}$ are $2.7\sigma$ ($1.5\%$)and $2.0\sigma$ ($0.1\%$) smaller, respectively.  The $\chi^2$ value for this fit is 536,738.04, equivalent to the \texttt{PHOENIX} model fits and marginally worse than the \texttt{ATLAS} models or the fitted limb-darkening coefficients.  This fit confirms the pattern suggested earlier, namely that stronger limb darkening leads to larger planet-star radius ratios and larger values for the inclination and $a/R_{\star}$.  If we consider the constraints imposed by the transit fits, stronger limb darkening means that for a grazing transit the planet must occult a relatively larger fraction of the star in order to produce the same apparent transit depth.  This effect will be even larger in visible light, and we conclude that accurate limb-darkening coefficients are essential when calculating the planet-star radius ratio and corresponding transmission spectrum for near-grazing transits.

It is difficult to diagnose the origin of the disagreement between \texttt{ATLAS} and \texttt{PHOENIX} stellar atmosphere models for GJ~436; \citet{kurucz05} note that the \texttt{ATLAS} models should not be used for stellar effective temperatures below 3500~K, as they do not include important low-temperature opacity sources such as TiO and VO.  However, these molecules primarily affect the star's visible and near-infrared spectra, and at 3500~K they are still relatively weak \citep{cushing05}.  Both disk-integrated and intensity spectra for the \texttt{ATLAS} models in this temperature range are featureless longward of 2.4~\micron, with the exception of the CO band between 4.3\,$-$\,5.0~\micron, whereas \texttt{PHOENIX} spectra show also clear molecular band structures, mainly due to H$_2$O and OH, between 2.5\,$-$\,3.6 $\mu$m and 6.5\,$-$\,8.0~\micron, with corresponding increases in the amount of limb darkening in these bands.  The presence of the CO band in both model sets would appear to explain the relatively good agreement in limb-darkening proÞles for the 4.5~\micron~Spitzer bandpass, but we were unable to determine the reason behind the missing mid-IR H$_2$O absorption features in the \texttt{ATLAS} models, which incorporate the strongest water lines \citep{cdrom26} from the Ames list of \citet{AmesH2O}.  It is possible that the spherical geometry used in the \texttt{PHOENIX} models (\texttt{ATLAS} models use a plane-parallel geometry) may also affect the resulting limb-darkening profiles \citep{orosz00,claret03}, but we find that \texttt{PHOENIX} models computed with a planet-parallel geometry show nearly identical limb darkening, with the exception of an exponential decline in the optically thin limb.  We conclude that the differing opacities in the 3.6 and 8.0~\micron~bands appear to be the most likely explanation for the disagreement between the limb-darkening profiles at these wavelengths.  In this case the change in the $\chi^2$ value indicates the differences between the two models are not statistically significant for this data set; near-IR grism spectroscopy of transits of GJ~436b, such as those obtained by \citet{pont08}, might help to better distinguish between these models.

\begin{deluxetable}{lrrrrcrrrrr}
\tabletypesize{\scriptsize}
\tablecaption{Global System Parameters \label{global_param}}
\tablewidth{0pt}
\tablehead{
\colhead{Parameter} & \colhead{\phantom{aaaaa}Value}}
\startdata
\emph{Transit Parameters} &\\
$i$(\degr) & $86.699^{+0.034}_{-0.030}$ \\
$a/R_{\star}$ & $14.138^{+0.093}_{-0.104}$ \\
$R_p/R_{\star}$\tablenotemark{a} & $0.08311\pm0.00026$ \\
Duration $T_{14}$ (d)\tablenotemark{b} & $0.04227\pm0.00016$\\
$T_{12}$~($\approx T_{34}$)\tablenotemark{b} (d)& $0.01044\pm0.00014$\\
$b$ & $0.8521\pm0.0021$\\
$a$ (A.U.)\tablenotemark{c} & $0.0287\pm0.0003$\\
$R_{\star}$ (\ensuremath{R_\odot})\tablenotemark{c} & $0.437\pm0.005$\\
$R_p$ (\ensuremath{R_\Earth})\tablenotemark{c} & $3.96\pm0.05$\\
$T_c$ (BJD) & $2454865.083208\pm0.000042$ \\
$P$ (d) & $2.6438979\pm0.0000003$ \\
&\\
\emph{Secondary Eclipse Parameters} &\\
8~\micron~depth & $0.0452\%\pm0.0027\%$ \\
$T_{\mbox{bright}}\tablenotemark{d}$ & $740\pm16$~K \\
Duration $T_{14}$ (d)\tablenotemark{e} & $0.04347$\\
$T_{12}$~($\approx T_{34}$)\tablenotemark{e} (d)& $0.00700$\\
Orbital phase & $0.58672\pm0.00017$\\
$e\cos(\omega)$ & $0.13775\pm0.00027$\\
$T_c(0)$ (BJD) & $2454866.63444\pm0.00082$ \\
$P$ (d) & $2.6438944\pm0.0000071$ \\
\enddata
\tablenotetext{a}{Calculated from the error-weighted average of the four 8~\micron~planet-star radius ratio; this value was used for secondary eclipse fits.}
\tablenotetext{b}{The transit duration $T_{14}$ is defined as the time from first to fourth contact (i.e., the start of ingress to the end of egress).  $T_{12}$ is the length of ingress, which is equal to the egress length in the limit of a circular orbit.  Our best-fit transit ingress and egress lengths differ by less than 3~s.}
\tablenotetext{c}{These parameters incorporate the stellar mass estimate of $0.452\pm0.013$~\ensuremath{M_\odot} from \citet{torres07}.}
\tablenotetext{d}{Brightness temperature is defined as the temperature required to match the observed planet-star flux ratio in the 8~\micron~\emph{Spitzer} band assuming that the planet radiates as a blackbody and using a \texttt{Phoenix} stellar atmosphere model ($T_{eff}=3585$~K, $\log(g)$=4.843) for the star.}
\tablenotetext{e}{The secondary eclipse duration and the length of ingress/egress were fixed in our fits.}
\end{deluxetable}

\subsubsection{Error Analysis}

We calculate uncertainties for our best-fit transit parameters using a Markov Chain Monte Carlo (MCMC) fit \citep[see, for example][]{ford05,winn07b} with a total of $6\times 10^6$ steps, fourteen independent chains, and 53 free parameters.  Free parameters in the fits include: $a/R_{\star}$, $i$, eight individual estimates of $R_P/R_{\star}$, eight transit times, eight constants, a linear function of time and linear and quadratic terms in $x$ and $y$ for each of the 3.6 and 4.5~\micron~transits (20 variables total), the amplitude $c_2$ and decay time $c_3$ from Eq. \ref{eq2} for the exponential fits to three 8.0~\micron~transits (6 variables total), and a linear function of time for the other 8.0~\micron~visit.  We assume a constant error for the points in each individual transit light curve, defined as the the uncertainty needed to produce a reduced $\chi^2$~equal to one for the best-fit transit solution.  

We initialize each chain at a position determined by randomly perturbing the best-fit parameters from our Levenberg-Marquardt minimization.  After running the chain, we search for the point where the $\chi^2$~value first falls below the median of all the $\chi^2$~values in the chain (i.e. where the code had first found the optimal fit), and discard all steps up to that point.  We calculate the uncertainty in each parameter as the symmetric range about the median containing 68\%~of the points in the distribution, except for the inclination and $a/R_{\star}$, which we allow to have asymmetric error bars spanning 34\% of the points above and below the median, respectively.  The distribution of values was very close to symmetric for all other parameters, and there did not appear to be any strong correlations between variables.  As a check we also carried out a residual permutation error analysis \citep{gillon07b,winn08}, which is sensitive to correlated noise in the light curve, on each individual transit.  At the start of each new permutation, we randomly drew values for the inclination and $a/R_{\star}$ from the simultaneous MCMC distribution and then fit for the corresponding best-fit values for the transit time and $R_p/R_{\star}$ in that step.  This ensures that our resulting error distributions for individual transit times and $R_p/R_{\star}$ values also take into account the uncertainties in the best-fit values for the inclination and $a/R_{\star}$.  In each case where both a MCMC and residual permutation uncertainty are available for a given parameter we use the higher of the two values.  We find that the MCMC fits generally produce larger uncertainties for the 8~\micron~observations, whereas for 3.6 and 4.5~\micron~data sets, which have higher levels of correlated noise, the residual permutation uncertainties are typically 50\% larger than the MCMC errors.

\subsubsection{Secondary Eclipse Fits}

We fit the secondary eclipses individually using the best-fit values for inclination and $a/R_{\star}$ from our transit fits but allowing the eclipse depths and times to vary freely.  Figure \ref{eclipse_phot_raw} shows the final binned data from these fits with the best fit normalizations for the detector ramp in each channel overplotted, and Figure \ref{eclipse_phot_norm} shows the binned data once these trends are removed, with best-fit eclipse curves overplotted.  Best-fit parameters for individual eclipses are given in Table \ref{eclipse_param}, and the error-weighted average (i.e., weights equal to $1/\sigma^2$) of these eclipse depths is reported in Table \ref{global_param}.  We find that fixing the time of eclipse to a constant value, defined here as the best-fit orbital phase, does not significantly change our best-fit eclipse depths, nor does it reduce the uncertainties in our measurement of those depths.  We calculate the uncertainties on individual eclipses using both a MCMC analysis and a residual permutation error analysis, again taking the higher of the two values as the final uncertainty for each parameter.

\section{Results}

\subsection{Orbital Ephemeris and Limits on Timing Variations}\label{timing}

\begin{figure}
\epsscale{1.2}
\plotone{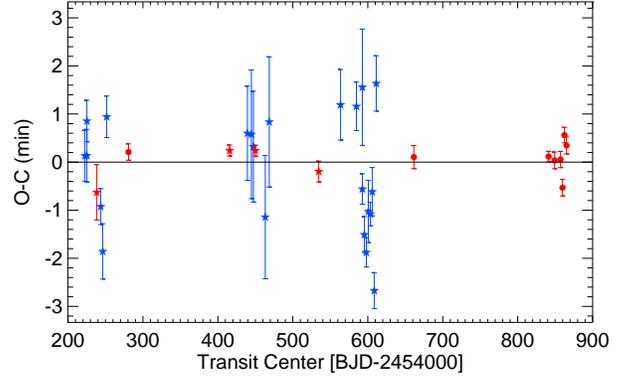}
\caption{Observed minus calculated transit times using the new best-fit ephemeris.  The dashed lines indicate the $\pm1\sigma$ uncertainty in the predicted transit times, with a solid line at $O-C$ equal to zero.  \emph{Spitzer} measurements from this paper are plotted as filled circles, and previously published observations are shown as filled stars.  The color of the points denotes the wavelength of the observations (blue for visible, red for IR).  Moving from left to right, transits between 200-300 BJD-2454000 are from \citet{shporer09} and \citet{caceres09}, transits between 400-500 BJD-2454000 with small uncertainties are from \citet{pont08a} and those with large uncertainties are from \citet{bean08}.  Between 530-620 BJD-2454000, observations are from \citet{coughlin08}, \citet{alonso08}, and \citet{ballard10a}.  Visible-light transit observations typically show larger timing variations than the IR observations, indicating that spot occultations may be responsible for the apparent timing variations.}
\label{transit_o_c}
\end{figure}

We fit the transit times given in Table \ref{transit_param}, together with the transit times published in \citet{pont08a,bean08,coughlin08,alonso08,shporer09,caceres09,ballard10a}, with the following equation,

\begin{align}\label{eq3}
T_c(n)= T_c(0)+n\times P
\end{align}
where $T_c$ is the predicted transit time as a function of the number of transits elapsed since $T_c(0)$ and $P$ is the orbital period.  We find that $T_c=2454865.083208\pm0.000042$ BJD and $P=2.6438979\pm0.0000003$~days.  As demonstrated by Fig, \ref{transit_o_c}, the 34 published transit times appear to be markedly inconsistent with a constant orbital period, with the most statistically significant outliers (6.2 and 7.1$\sigma$, respectively), occurring during the sequence of eight transits observed by the EPOXI mission between UT 2008 May 5-29 \citep{ballard10a}.  The most significant deviations in the \emph{Spitzer} transit data presented here occur during the last three visits (UT 2009 Jan 28 - Feb 2), and range between -3.1 and +3.5$\sigma$ in significance.  Given the size of these discrepancies, it is perhaps not surprising that the reduced $\chi^2$ value for the linear fit to Eq. \ref{eq3} is 6.8 (total $\chi^2$ of 216.4, 34 points, two degrees of freedom).  It is unlikely that the observed deviations could be explained by perturbations from a previously unknown second planet in the system, as the measured transit times shift by as much as several minutes on time scales of only a few days (i.e., a single planet orbit).  As we discuss in more detail in \S\ref{star_spots}, we believe the presence of occulted star spots in a subset of the transit light curves is the most likely explanation for the observed deviations.

\begin{figure}
\epsscale{1.2}
\plotone{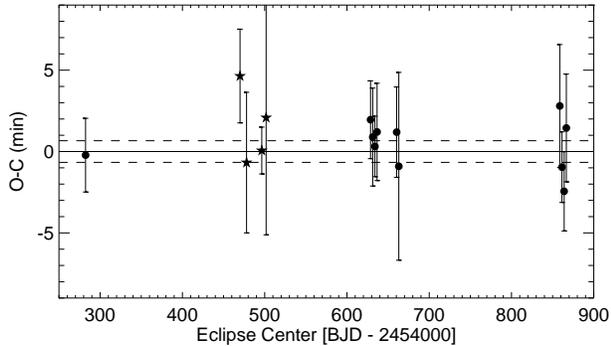}
\caption{Observed minus calculated secondary eclipse times using the best-fit period from the transit fits and allowing the phase of the secondary eclipse to vary freely.  Filled circles are eclipse times reported in this paper, and filled stars are additional 3.6 and 4.5~\micron~eclipse times from \citet{stevenson10}.  The solid line indicates the best-fit phase, with $\pm1\sigma$ uncertainties plotted as dashed lines.}
\label{eclipse_o_c}
\end{figure}

We carry out a similar fit to the secondary eclipse times given in Table \ref{eclipse_param}, along with the additional secondary eclipse times reported in \citet{stevenson10}, and find that $T_c(0)= 2454866.63444\pm0.00082$~BJD and $P=2.6438944\pm0.0000071$~days.  This period is consistent with the best-fit transit period to better than $1\sigma$, and we therefore conclude that there is no evidence for orbital precession in this system.  We also see no evidence for statistically significant variations in the secondary eclipse times (see Fig. \ref{eclipse_o_c}), as would be expected if the shifted transit times were due to occulted spots, but our measurements are not precise enough to rule out timing variations of the same magnitude as those observed in the transit data.  If we fix the orbital period to the value from the transit fits and subtract the 28~s light travel time delay for this system \citep{loeb05}, we find that the secondary eclipses occur at an orbital phase of $0.58672\pm0.00017$, consistent with the best-fit phase from \citet{stevenson10}.

We can use the offset in the best-fit secondary eclipse time to calculate a new estimate for $e\cos(\omega)$.  We find that the secondary eclipse occurs $330.18\pm0.67$ minutes later on average than the predicted time for a circular orbit, including the correction for the light travel time.  We can convert this to $e\cos(\omega)$ using the expression reported in Eq. 19 of \citet{pal10}.  Note that this expression is more accurate than the commonly used approximation of $e\cos(\omega)\approx \frac{\pi \delta t}{2P}$ \citep[e.g.,][]{charbonneau05,deming05}, where $\delta t$ is the delay in the measured secondary eclipse time and $P$ is the planet's orbital period.  We find that using the less accurate approximation gives $e\cos(\omega)=0.13622\pm0.00026$, while the equation from P\'al et al. yields $e\cos(\omega)=0.13754\pm0.00027$, a $4\sigma$ difference in this case \citep[see also,][]{sterne40,dekort54}.  If we take the best-fit longitude of pericenter from the radial velocity fits, $334\degr \pm 10\degr$, we find an orbital eccentricity equal to $0.153\pm0.014$.  This is consistent with the current best-fit orbital eccentricity from radial velocity data alone, $e=0.145\pm0.017$ (A. Howard, personal communication, 2010).

\begin{deluxetable}{lrrrrcrrrrr}
\tabletypesize{\scriptsize}
\tablecaption{$a/R*$ and Inclination Values from Independent Transit Fits\label{free_transit_fits}}
\tablewidth{0pt}
\tablehead{
\colhead{UT Date} & \colhead{$\lambda$~(\micron)} & \colhead{\phantom{aaaa}$i$ (\degr)} & \colhead{$a/R_{\star}$}}
\startdata
UT 2007 Jun 29 & 8.0 & $86.68 \pm0.12$ & $14.11\pm0.35$ \\
UT 2008 Jul 14 & 8.0 & $86.54 \pm0.12$ & $13.67\pm0.36$ \\
UT 2009 Jan 9 & 3.6 & $86.76 \pm0.07$ & $14.40\pm0.22$ \\
UT 2009 Jan 17 & 4.5 & $86.85 \pm0.10$ & $14.60\pm0.32$ \\
UT 2009 Jan 25 & 8.0 & $86.70 \pm0.14$ & $14.19\pm0.43$ \\
UT 2009 Jan 28 & 3.6 & $86.67 \pm0.07$ & $13.96\pm0.20$ \\
UT 2009 Jan 30 & 4.5 &$86.58 \pm0.10$ & $13.76\pm0.28$ \\
UT 2009 Feb 2 & 8.0 & $86.80 \pm0.14$ & $14.49\pm0.44$ \\
\enddata
\end{deluxetable}

\subsection{System Parameters from Transit Fits}

In this work we examine two transits obtained at 3.6~\micron, two transits at 4.5~\micron, and four transits at 8.0~\micron.  We carry out two sets of transit fits, one where the ratio of the orbital semi-major axis to the stellar radius $a/R_{\star}$ and the orbital inclination $i$ are allowed to vary freely, and the other where they have a single common value for all visits.  In all cases we allow the planet-star radius ratio $R_p/R_{\star}$ and best-fit transit times to vary independently for each visit.  In fits where $a/R_{\star}$ and $i$ are allowed to vary individually we find no evidence for statistically significant variations in either of these parameters (see Table \ref{free_transit_fits}), and we therefore proceed assuming that these parameters have a single common value in our subsequent analysis.  Our best-fit values for $i$, $a/R_{\star}$, and $R_p/R_{\star}$ are consistent with those reported by \citet{ballard10a} to better than $1\sigma$, and the impact parameter $b$ and transit duration $T=T_{14}-T_{12}=0.0318\pm0.0007$ days that we derive from our fits are similarly consistent with the value reported by \citet{pont08}.

Although the best-fit orbital inclination and $a/R_{\star}$ appear to be consistent with a constant value over the approximately two year period spanned by our observations, we do see evidence for statistically significant differences in the transit depths \emph{within the same Spitzer bandpass} (see Fig. \ref{depth_comparison}).  We would expect to see the transit depth vary with wavelength due to absorption from the planet's atmosphere, but this signal should remain constant from epoch to epoch for observations in the same bandpass.  If we compare individual visits in a given bandpass, we find that the two 3.6~\micron~radius ratios, measured on UT 2009 Jan 9 and 28, are inconsistent at the $4.7\sigma$~level.  The two 4.5~\micron~radius ratios, measured on UT 2009 Jan 17 and 30, differ by $2.9\sigma$.  The four 8~\micron~transits, measured on UT 2007 Jun 29, UT 2008 Jul 14, UT 2009 Jan 28, and UT 2009 Feb 2, differ from the error-weighted average by $0.2\sigma$, $1.0\sigma$, $1.5\sigma$, and $2.0\sigma$, respectively.  These offsets are still present in the fits where the inclination and $a/R_{\star}$ are allowed to vary individually, indicating that the discrepancy cannot be due to a change in these two parameters.  

\section{Discussion}

\subsection{Transit Depth Variations}

In the sections below we consider three possible explanations for the observed depth variations: first, that the effective radius of the planet is varying in time, second, that residual correlated noise in the data affected the best-fit transit solutions, and third, that spots or other stellar activity produced apparent depth variations.

\subsubsection{A Time-Varying Radius for the Planet}\label{time_var_radius}

We first consider the possibility that the radius of the planet is changing in time, either due to thermal expansion of the atmosphere or the presence of a variable cloud layer at sub-mbar pressures.  We require a change in radius of approximately 4\% in order to match both of the measured 3.6~\micron~transit depths; if this change is due to thermal expansion, we can estimate the energy input required using simple scale arguments.  

The effective change in the planet's radius due to heating of the atmosphere depends on both the amount of heating and the range in pressures over which this heating takes place.  We use the secondary eclipse depths in \S\ref{dayside_var} to place an upper limit on the allowed change in temperature at the level of the mid-IR photosphere, and then calculate the corresponding range in pressure that must be heated by this amount in order to increase the radius of the planet by 4\%.  If we assume a hydrogen atmosphere with a baseline temperature of 700~K, we find a corresponding scale height of approximately 240~km, where the scale height is defined as $H=\frac{kT}{\mu g}$, $T$ is the temperature of the atmosphere, $\mu$ is the mean molecular weight, and $g$ is the surface gravity.  We know from the secondary eclipse observations described in \S\ref{dayside_var} that the temperature of the planet's dayside atmosphere must change by less than 30\%, which would correspond to an upper limit of 100~km on corresponding changes in the planet's scale height.

\begin{figure}
\epsscale{1.6}
\plotone{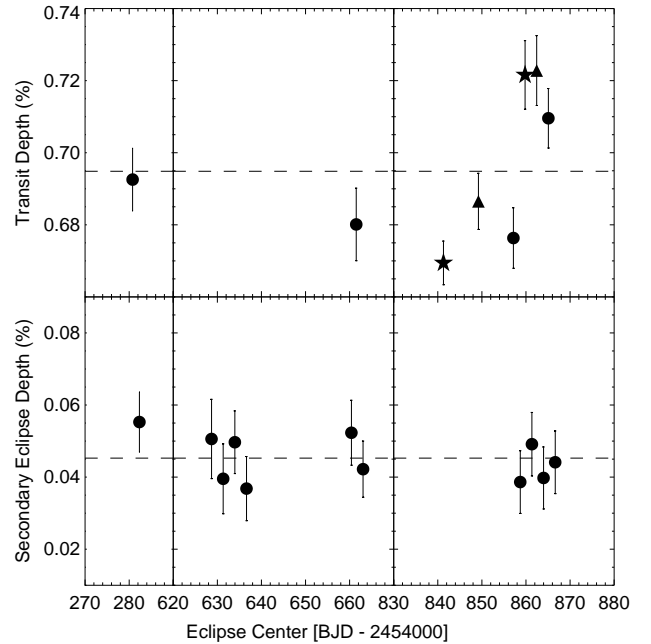}
\caption{Best-fit transit (upper panel) and secondary eclipse (lower panel) depths as a function of time.  Average transit/eclipse depths are shown as a dashed line in each plot.  3.6~\micron~observations are denoted with stars, 4.5~\micron~observations are denoted with triangles, and 8.0~\micron~observations are marked with solid circles.  The most recent three transits are systematically high when compared to earlier transit visits; this appears to coincide with a decrease in the total visible light flux from the star (Fig. \ref{stellar_rot}), suggesting that the fractional spot coverage on the star's visible face was increasing in time during the later part of our observations.}
\label{depth_comparison}
\end{figure}

In order to calculate the required energy input to produce the observed change in radius, we must first determine the range of pressures affected by this heating.  We model the planet as an interior region with a constant temperature, surrounded by an outer envelope that expands and contracts freely with changing temperature.  We set the upper boundary on this region equal to 50 mbar, corresponding to the approximate location of the $\tau=1$ surface in the mid-infrared.  As illustrated in Fig. \ref{atm_models}, opaque clouds at this pressure suppress but do not entirely remove absorption features in the planet's transmission spectrum at these wavelengths, making this a reasonable estimate for the location of the $\tau=1$ surface.  We assume that when the planet is heated the scale height changes by 100~km, which requires the lower boundary of the heated region to be located at a pressure of approximately 1 bar in order to produce a 1\% expansion in radius.  If we then calculate the change in the planet's gravitational energy corresponding to this expansion, we find that an energy input of approximately $10^{26}$~J is required.  Repeating this calculation for a 4\% increase in radius, we find a lower boundary at 8,000 bars and a corresponding energy input of $10^{30}$~J.  The insolation received by the planet is $10^{20}$~W, which gives an energy budget of $10^{25}$~J per orbit.  When we examine Fig. \ref{depth_comparison} we find that that the observed change in radius occurs primarily between the third and fourth visits (UT 2009 Jan 25-28).  This would require an energy input as much as $10^5$ times higher than the total insolation over this epoch, which is clearly unphysical.

One alternative explanation for the observed change in radius would be to invoke the presence of intermittent, high-altitude clouds.  Such clouds could produce a change in the apparent radius of the planet across multiple bands without requiring any actual heating or cooling of the atmosphere.  In this picture, smaller radii for the planet would correspond to the cloud-free state, while larger radii would require the presence of an additional cloud layer.  A change of 4\% in apparent radius would require the clouds to form at a pressure approximately 100 times lower than the location of the nominal cloud-free radius.  In \S\ref{modeling_disc} we find that the average pressure of the $\tau=1$ surface for the nominal methane-poor (green) model between $3-10$~\micron~is 40 mbar, indicating that the clouds would have to extend to 0.4 mbar to explain the largest measured 3.6~\micron~radius for the planet.  This conclusion is reasonably independent of our assumed composition, as the average $\tau=1$ surface for the methane-rich (blue) model is located at 30 mbar.   Gravitational settling would presumably pose a challenge for cloud layers at sub-mbar levels, but vigorous updrafting of condensate particles might compensate for this effect.  The broadband nature of the data presented here make it difficult to directly test this hypothesis; we therefore recommend the acquisition of high signal-to-noise, near-infrared grism spectroscopy over multiple transits in order to resolve this issue.  A 0.5 mbar cloud layer would lead to a near-featureless transmission spectrum whereas a lower cloud layer would still exhibit many of the same absorption features as a cloud-free atmosphere.  Such a data set would also allow us to test the theory, outlined in \S\ref{star_spots}, that the observed transit depth variations are due to the occultation of regions of non-uniform brightness on the surface of the star, as these regions should also produce a wavelength-dependent effect.

\subsubsection{Poorly Corrected Systematics}\label{det_norm}

It is possible that poorly corrected instrument effects, such as the intrapixel sensitivity variations at 3.6 and 4.5~micron, or the detector ramp at 8.0~\micron, might lead to variations in the measured transit depth.  Because there is complete overlap between the positions spanned by the star in the in-eclipse and out-of-eclipse data for all 3.6 and 4.5~\micron~visits, fits that inadequately describe the pixel response as a function of position should fail equally for both sections of the light curve.  The UT 2009 Jan 30 transit serves as an example of imperfectly removed detector effects, as the residuals display a sawtooth signal with a shape and timescale similar to the original intrapixel sensitivity variations (see \S\ref{short_phot} for a more detailed discussion of this light curve).  Conversely, it is much more difficult to explain the 3.6 micron transit on UT 2009 Jan. 28 with this scenario, as there appears to be a large dip in residuals during ingress, but when the star spans the same pixels in the out-of-eclipse data we see no comparable deviations.  

In this section we consider an alternate decorrelation function that better accounts for small-scale variations in the intrapixel sensitivity function as discussed in \citet{ballard10}.  Following the discussion in Ballard et al., we describe the intrapixel sensitivity variations using a position-weighted average of the time series after the best-fit transit function and linear function of time from the position fits described above has been divided out.  Unlike Eq. \ref{eq1}, this formalism does not assume a functional form for the intrapixel sensitivity variations and therefore should in principle produce an unbiased correction for these variations.  We calculate the weighting function as follows: 

 \begin{align}\label{eq4}
W(x_i,y_i)=\frac{\sum\limits_{i\neq j}{\exp\left(-\frac{(x_j-x_i)^2}{2\sigma _x^2}\right)\exp\left(-\frac{(y_j-y_i)^2}{2\sigma _y^2}\right)f_j}}{\sum\limits_{i\neq j}{\exp\left(-\frac{(x_j-x_i)^2}{2\sigma _x^2}\right)\exp\left(-\frac{(y_j-y_i)^2}{2\sigma _y^2}\right)}}
\end{align}
where $x_i$ and $y_i$ are the $x$ and $y$ positions of the $i^{th}$ frame, $x_j$ and $y_j$ are the $x$ and $y$ positions for the rest of the time series.  We optimize our choice of $\sigma _x$ and $\sigma_y$ to produce the smallest possible scatter in the final time series when we fix the transit light curve to the best-fit solutions listed in Table \ref{transit_param}.  We find that the preferred values range between $0.0053-0.0120$ pixels in $\sigma _x$ and $0.0024-0.0045$ pixels in $\sigma_y$ for the four 3.6 and 4.5~\micron~transits examined here.  For ease of computation we bin our time series in intervals corresponding to one point per original set of 64 images (in some instances there are less than 64 images in a given bin after removing outliers) and iteratively calculate the weighting function and the linear function of time plus transit fits until we converge to a consistent solution.  

Once we have a final solution we calculate the weighting function for the unbinned data and carry out a final fit for the transit function to determine our best-fit transit depth.  In this case we fix the inclination and $a/R_{\star}$ to their best-fit values from the simultaneous fits to all transits described in \S\ref{transits}, which allows us to fit each transit individually using the weighting function while still preserving the constraints imposed in a simultaneous fit.  We find that in all cases we obtain transit depths and times that are consistent with the values from our fits using Equation \ref{eq1}, with a standard deviation that is comparable or slightly worse than that achieved with our polynomial fits.  

We also carried out a second set of fits in which we derived our corrections for the intrapixel sensitivity variations using only the out-of-transit data, and found that our best-fit planet-star radius ratios changed by less than $0.4\sigma$ in all cases.  Because the star samples the same regions of the pixel in both the in-transit and out-of-transit data, it is possible to obtain an equivalently good correction for the intrapixel sensitivity variations using only the out-of-transit points.  Conversely, this means that poor corrections for this effect should produce equally large deviations in both the in-transit and out-of-transit regions of the light curve.  As we will discuss in the following section, we find that the residuals for the deepest transits in these two bands have a significantly higher RMS in transit than out of transit.  This behavior is inconsistent with our expectations for poorly corrected instrument effects, and we therefore conclude that it is unlikely that these effects are responsible for the discrepant transit depths measured at 3.6 and 4.5~\micron.  

At 8.0~\micron~we fit the data with a single or double exponential function to describe the smoothly varying detector ramp.  In \citet{agol10} we conclude that this functional form avoids correlations between the slope of the ramp and the measured transit or eclipse depth; however we check this assertion using our 8~\micron~data as well.  For our 8~\micron~transit fits we find that the exponential term has a coefficient of [0.00156, 0.00000, 0.00288, 0.00299], corresponding to planet-star radius ratios of [0.08234, 0.08162, 0.08138, 0.08336], where we have set the amplitude of the exponential term to zero for the transit occurring in the middle of our 70-hour phase curve observation.  For the eleven secondary eclipse observations we find coefficients of [0.00645, 0.00627, 0.00194, 0.00433, 0.00534, 0.00140, 0.00000, 0.00359, 0.00321, 0.00353, 0.00262], corresponding to eclipse depths of [0.0552, 0.0507, 0.0395, 0.0495, 0.0367, 0.0523, 0.0421, 0.0386, 0.0491, 0.0397, 0.0441], respectively, where we have set the exponential coefficient to zero for the secondary eclipse at the end of the phase curve observation.  We find no evidence for any correlation between the slope of the exponential function and the measured transit or secondary eclipse depths.  As an additional check we also confirm that there is no correlation between these depths and either the measured sky background or the total stellar flux given in Table \ref{obs_table}.

\begin{figure}
\epsscale{1.2}
\plotone{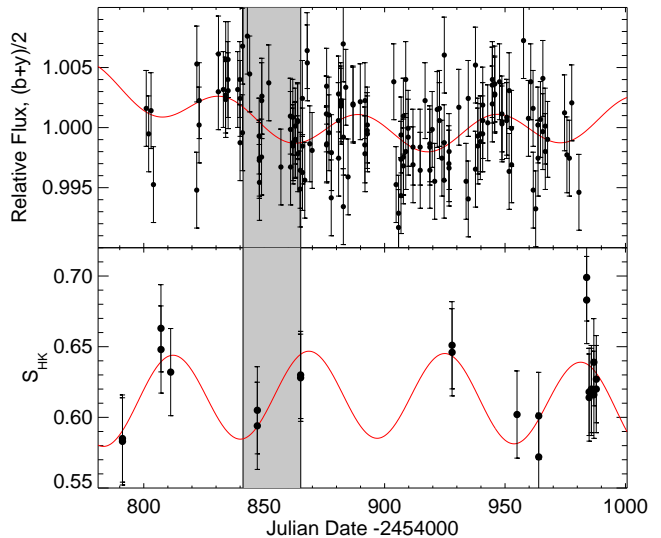}
\caption{The upper panel plots APT measurements of variation in the averaged Stromgren $b$+$y$ band fluxes (filled circles) obtained for GJ 436 between UT 2008 Nov 30 and UT 2009 May 29, where the epoch of our 2009 \emph{Spitzer} observations is denoted by the grey shaded region.  These bandpasses are sensitive to the rotation-modulated flux of the star, which we find has a best-fit period of 57 days during this epoch.  We overplot a red curve showing our best sine-curve fit for the spot modulation, together with a quadratic function of time to describe the evolution of the spot coverage on longer time scales.  The lower panel shows the measured \sval~values for GJ~436 from the Keck HIRES instrument during this same period \citep{isaacson10}, with a best-fit sine + quadratic function overplotted in red.  Error bars for both panels are set equal to the standard deviation of the residuals.  Although the best-fit period of 57 days for the \sval~data during this epoch is only marginally significant, the \sval~values appear to be anti-correlated with the flux variations.  This is consistent with a model in which increased magnetic activity is associated with the presence of spots or other dark regions on the surface of the star.}
\label{stellar_rot}
\end{figure}

\begin{figure}
\epsscale{1.2}
\plotone{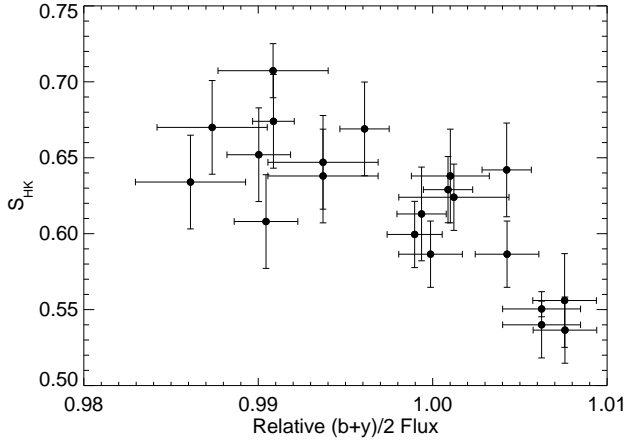}
\caption{Averaged Stromgren $b$ band fluxes vs. \sval~for GJ~436 over a period of seven years.  Because the sampling for the \sval~measurements is much lower than for the flux measurements in most seasons, for the purposes of this plot the $b$ fluxes are defined as the average of all $b$ band measurements taken within one day of the \sval~measurement.  We also average \sval~values taken on the same night, scaling the flux and \sval~error bars from Fig. \ref{stellar_rot} by the square root of the number of measurements in each bin.  We find that using $b$ photometry instead of the averaged $b$+$y$ fluxes results in increased scatter but also strengthens the observed correlation.}
\label{bflux_vs_sval}
\end{figure}

\subsubsection{Stellar Variability}\label{star_spots}

The presence of spots or faculae on the visible face of the star can have two distinct effects on the measured light curve for a transiting planet.  Non-occulted spots on the visible face of the star reduce the star's total flux, increasing the measured transit depth, while spots occulted by the planet cause a small positive deviation in the light curve with a time scale proportional to the physical size of the occulted spot \citep[e.g., ][]{rabus09}, and occulted faculae would have the opposite effect.  The early K dwarf HD 189733 ($T_\mathrm{eff}= 5100$~K) is perhaps the best-studied example of an active star with a transiting hot Jupiter \citep[e.g.,][]{bakos06,pont08a,desert10}, but the late G dwarf CoRoT-2 ($T_\mathrm{eff}=5600$~K) also exhibits a high level of spot activity that may have resulted in early overestimates of its planet's inflated radius \citep{guillot10}.  This problem is likely to be even more common for M dwarfs, and in fact several instances of occulted spots were reported in transit light curves for the super-Earth GJ~1214b, which orbits a 3000~K primary \citep{carter10,berta10,kundurthy10}.  

Although it is important to correct for these effects in any transit fit, it is particularly crucial when comparing non-simultaneous, multi-wavelength transit observations such as the ones described in this paper, which require a relative precision of better than a part in $10^{-4}$ in the measured transit depths.  We evaluate the likely impact of GJ~436's activity on the measured transit depths using several complementary approaches.  First, we estimate the average activity level on GJ~436 by measuring the amount of emission in the cores of the Ca II H \& K lines; in \citet{knutson10} we determined that GJ 436 had an average \sval~of 0.620.  \citet{isaacson10} found that other stars in the California Planet Search database with similar $B-V$ colors have \sval~values ranging between $0.5-2.0$, indicating that GJ~436b is relatively quiet for its spectral type.  \citet{demory07} report that this star's rotation period is greater than 40 days, consistent with upper limits on \vsini~of 1 km/s \citep{jenkins09}, also suggesting that it is likely to be relatively old and correspondingly quiet.  The upper limit of 3 km/s on \vsini~from spectroscopy \citep{butler04} is also consistent with an inclined or pole-on viewing geometry, although it is not required as long as the star's rotation period is longer than 7 days. 

Rather than relying on these indirect measures of activity, we can also directly measure the amplitude of the star's rotation-modulated flux variations using visible-light ground-based observations.  We obtained observations of GJ~436 in Str\"omgren \emph{b} and \emph{y} filters over a span of approximately six months surrounding our 2009~\emph{Spitzer} transit and secondary eclipse observations from an ongoing monitoring program carried out with the T12 0.8 m APT at Fairborn Observatory in southern Arizona \citep{henry99,eat03,henry08}.  In these observations the telescope nodded between GJ~436 and three comparison stars of comparable or greater brightness, which were then used to correct for the effects of variable seeing and airmass.  We find that during the period between UT 2009 Jan 9 - Feb 4, when a majority of our transit data was obtained, the star varied in flux by less than a few mmag in visible light (Figure \ref{stellar_rot}).  We carry out a similar check for variability in the infrared using the fifteen 8~\micron~flux estimates listed in Table \ref{obs_table}, which we find have a standard deviation of 0.07\%.  Both of these measurements indicate that the star is very nearly constant in flux in both visible and infrared light, and we can therefore rule out non-occulted spots as the cause of the observed transit depth variations.

We also use these same data to search for periodicities corresponding to GJ~436b's rotation period.  If we fit the combined $b$ and $y$ band fluxes with a sine function plus a quadratic function of time as shown in Fig. \ref{stellar_rot}, we find a best-fit period of 56.5 days.  We calculate a Lomb-Scargle periodogram \citep{lomb76,scargle82} for these data and find that this period has a false alarm probability of only 2\%, which we determine using a bootstrap Monte Carlo analysis.  We find a nearly identical best-fit period of 56.6 days in the \sval~values measured with Keck HIRES during this epoch \citep{isaacson10}, but with a much higher false alarm probability of approximately 20\%.  We also examine the correlation between the measured $b$ fluxes and \sval~values over the six-year period in which both were available (Fig. \ref{bflux_vs_sval}), and find that these parameters are negatively correlated.  Taken together, these data indicate that the small observed variations in GJ~436's visible-light fluxes are likely connected with the presence of regions of increased magnetic activity on the visible face of the star.    

Although such low-amplitude flux variations generally indicate that a star has relatively few spots, there are two important exceptions.  First, if the spots are uniformly distributed in longitude, it is theoretically possible to have a star with significant spot coverage and an effectively constant flux.  It would not be surprising if the occurrence rate and distribution of spots was different for M stars than for G or K stars, but in GJ 436b's case the lack of any flux variations larger than a few mmag would seem to place a strong limit on the allowed spot distributions.   We can quantify this limit if we assume that the deviation of approximately 0.08\% in the first part of the 3.6~\micron~transit light curve from UT 2009 Jan 28 shown in Fig. \ref{transit_phot_norm} is due to the occultation of a bright region on the star.  This region must have a surface intensity that is 12\% brighter than the rest of the star in order to produce the observed deviation.  If we compare \texttt{PHOENIX} models with varying effective temperatures integrated over this band, we find that the star's temperature must increase by approximately 200~K in the affected region in order to match this surface intensity.  We know that the total rotational modulation in the star's visible-light flux must remain below 0.1\%, and we estimate that an increase of 12\% in the 3.6~\micron~surface intensity should produce an increase of approximately 65\% in the Str\"omgren (\emph{b}+\emph{y})/2 band.  In this case the fractional area covered by active regions on the star must vary by less than 0.15\% from the most active to the least active hemisphere.  Of course, it is possible that the stellar atmosphere models do not provide an accurate match for the spectra of these active regions; if we instead use the measured 3.6~\micron~flux contrast of 12\%, we find a more conservative limit of 1\% on variations in the area affected by stellar activity.

A second, more plausible scenario involves tilting the rotation axis of star so that we are viewing it closer to pole-on, which would effectively suppress the amplitude of rotational flux variations regardless of spot coverage.  If we assume that the star's spin axis is randomly oriented with respect to our line of sight, the probability that it will fall within $45\degr$ of a pole-on view is 30\%.  In this scenario the star could be highly spotted, allowing for frequent occultations of spots by the planet, while still displaying a small rotational flux modulation.  This scenario would require the planet's orbit to be misaligned with respect to the star's rotation axis, but such misalignments are commonly seen in other transiting planet systems \citep{winn10a}.   Although the Rossiter-McLaughlin effect has never been successfully measured for GJ~436b, \citet{winn10b} find that the Neptune-mass planet HAT-P-11b, which is perhaps the best analogue to the GJ~436 system, has a sky-projected obliquity of $103\degr^{+26\degr}_{-10\degr}$ indicating that this system is significantly misaligned.  If most close-in planets start out misaligned and are then gradually brought into alignment through tidal interactions with their host star as proposed by \citet{winn10a}, the fact that HAT-P-11b still maintains both a non-zero orbital eccentricity and a significant misalignment would seem to suggest that the same could also be true for GJ~436b.    

If we proceed with the hypothesis that GJ~436 is both spotty and tilted with respect to our line of sight, we can then search for evidence of occulted spots in the light curves with discrepant transit depths.  We first compare the relative standard deviations of the in-transit ($\sigma_\texttt{in}$) and out-of-transit ($\sigma_\texttt{in}$) residuals plotted in Fig. \ref{transit_phot_norm}:

\begin{align}\label{eq5}
\sigma_\texttt{rel} = \frac{\sigma_\texttt{in} - \sigma_\texttt{out}}{\sigma_\texttt{out}}
\end{align}
We list the measured values of $\sigma_\texttt{rel}$ for all eight transit observations in Table \ref{sigma_table}.  Both the 3.6~\micron~transit on UT 2009 Jan 28 and the 4.5~\micron~transit on UT 2009 Jan 30 appear to have inflated values of $\sigma_\texttt{rel}$, as would be expected if the planet occulted active regions on the star during these visits.  We can quantify the statistical significance of the measured $\sigma_\texttt{rel}$ values if we assume that both the in-transit and out-of-transit points are drawn from the same underlying Gaussian distribution, and then ask how many times in a sample of 100,000 random trials we measure a value of $\sigma_\texttt{rel}$ greater than or equal to the value calculated directly from our observations.  In each trial we generate two synthetic data sets, each with the appropriate length corresponding to either the in-transit or out-of-transit measurements, and then calculate the standard deviation of each distribution and the corresponding value of $\sigma_\texttt{rel}$.  In the 3.6 micron transit observation on UT 2009 Jan 9 there are 81,848 out-of-transit flux measurements and 25,482 in-transit flux measurements, and we find that over 100,000 trials, we obtain a value of $\sigma_\texttt{rel}$ greater than or equal to the measured value of 0.2\% approximately 36\% of the time.  Repeating the same calculation for the 3.6 micron transit observed on Ut 2009 Jan 28, which has 82,238 out-of-transit points and 25,530 in-transit points, we obtain $\sigma_\texttt{rel}$ greater than or equal to the measured value of 1.4\% only 0.23\% of the time.  We list the corresponding probabilities for all eight transits in Table \ref{sigma_table}.

\begin{deluxetable}{lrrrrcrrrrr}
\tabletypesize{\scriptsize}
\tablecaption{A Comparison of the In-Transit vs. Out-of-Transit Standard Deviations \label{sigma_table}}
\tablewidth{0pt}
\tablehead{
\colhead{UT Date} & \colhead{$\lambda$~(\micron)} &  \colhead{$N_\texttt{in}$\tablenotemark{a}} & \colhead{$N_\texttt{out}$\tablenotemark{a}} & \colhead{$\sigma_\texttt{rel}$} & \colhead{$P(\sigma_\texttt{rel})$ \tablenotemark{b}} }
\startdata
\emph{Unbinned Data} & & & & & \\
UT 2007 Jun 29 & 8.0 & 7,924 & 17,956 & -1.3\% & 0.92 \\
UT 2008 Jul 14 & 8.0 & 7,895 & 25,012 & +1.4\% & 0.059 \\
UT 2009 Jan 9 & 3.6 & 25,482 & 81,848 & +0.2\% & 0.36 \\
UT 2009 Jan 17 & 4.5 & 25,954 & 82,212 & -0.3\% & 0.70 \\
UT 2009 Jan 25 & 8.0 & 7,910 & 25,334 & +0.1\% & 0.44 \\
UT 2009 Jan 28 & 3.6 & 25,536 & 82,238 & +1.4\% & 0.0023 \\
UT 2009 Jan 30 & 4.5 & 25,955 & 62,334 & +1.1\% & 0.018 \\
UT 2009 Feb 2 & 8.0 & 7,890 & 25,318 & +0.1\% & 0.45 \\
\emph{Binned Data} & & & & & \\
UT 2007 Jun 29 & 8.0 & 126 & 287 & -13.6\% & 0.97 \\
UT 2008 Jul 14 & 8.0 & 126 & 399 & -4.9\% & 0.75 \\
UT 2009 Jan 9 & 3.6 & 411 & 1311 & +3.3\% & 0.21 \\
UT 2009 Jan 17 & 4.5 & 412 & 1310 & -3.4\% & 0.80 \\
UT 2009 Jan 25 & 8.0 & 126 & 403 & +9.7\% & 0.093 \\
UT 2009 Jan 28 & 3.6 & 411 & 1311 & +37.5\% & $1\times 10^{-6}$ \\
UT 2009 Jan 30 & 4.5 & 412 & 989 & -1.4\% & 0.63 \\
UT 2009 Feb 2 & 8.0 & 126 & 403 & +2.9\% & 0.34 \\
\enddata
\tablenotetext{a}{Number of in-transit and out-of-transit points.}
\tablenotetext{b}{Probability that the standard deviation of the in-transit data would be greater than the standard deviation of the out of transit data by an amount $\sigma_\texttt{rel}$ if both data sets are drawn from the same underlying Gaussian distribution.}
\end{deluxetable}

We also repeat this same test with data that has been binned in sets of 64 images, corresponding to 10 s bins at 3.6 and 4.5 um and 30 s bins at 8 um.  This allows us to evaluate the relative contribution that correlated noise makes to the in-transit and out-of-transit variances, as the photon noise should be be reduced by a factor of 8 in these bins (also see Fig. 3).  In this case we carry out 1,000,000 random trials for each visit, as each simulated data set is much smaller and the computations are correspondingly fast.  We find that for the binned Jan 9 light curve there are 1311 points out of eclipse and 411 points in eclipse.  In this case $\sigma_\texttt{rel}$ is 3.3\%, and we obtain values greater than or equal to this number in 21\% of our random trials.  Repeating this calculation for the UT 2009 Jan 28 visit, we find that the measured value of $\sigma_\texttt{rel}$ is 37\% (i.e., a standard deviation that is 37\% higher in eclipse than it is out of eclipse), with 1311 points out of eclipse and 411 points in eclipse.  In our simulations assuming a single Gaussian probability distribution for both segments, this level of disagreement occurred only once in $10^6$ trials.  We find that in all other visits, including the 4.5 micron transit observed on UT 2009 Jan 30, the binned data in and out of eclipse are consistent with a single distribution. 

\begin{figure*}
\epsscale{1.2}
\plotone{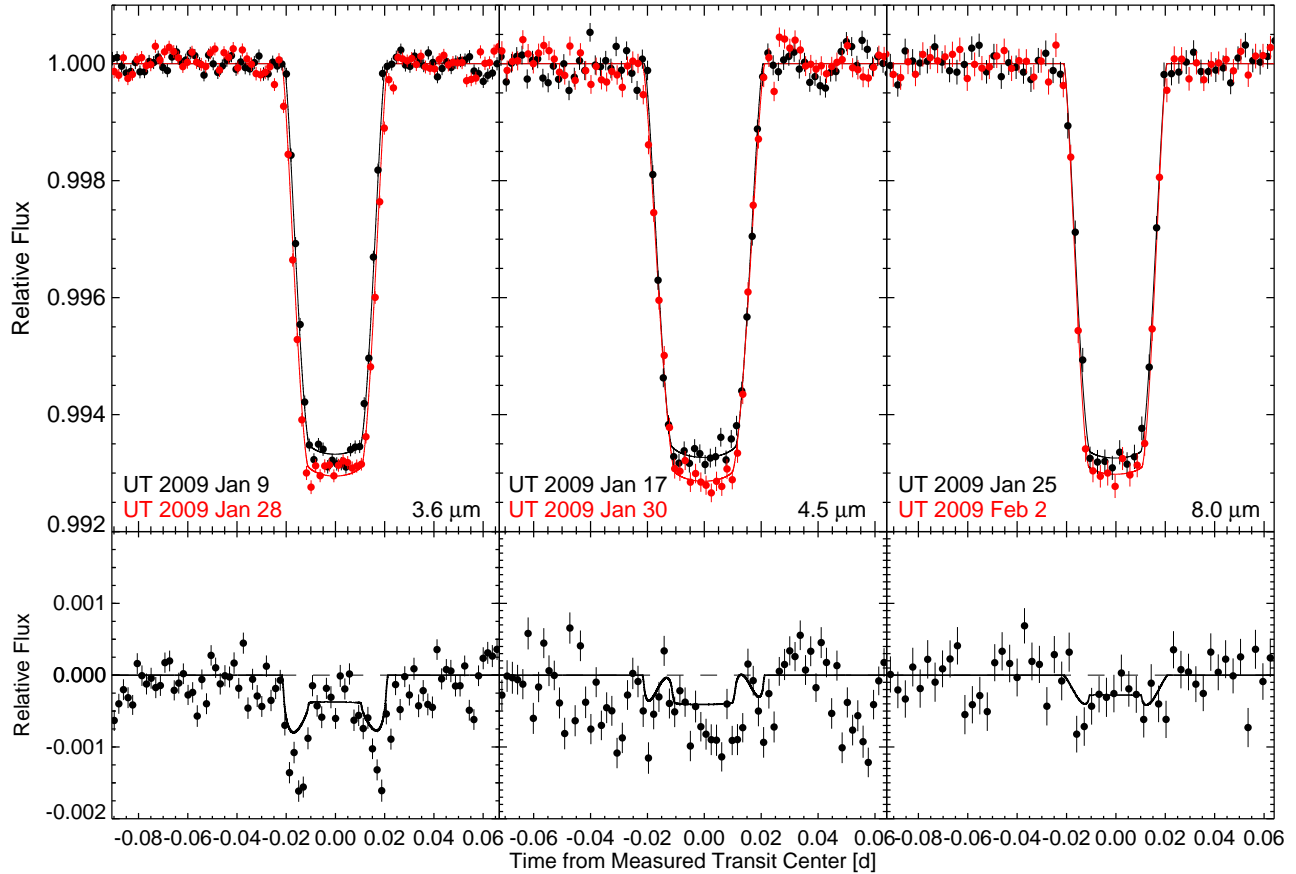}
\caption{This plot shows the six transits observed in Jan/Feb. 2009.  The left panel shows both 3.6~\micron~transits, the middle panel shows both 4.5~\micron~transits, and the right panel shows both 8.0~\micron~transits.  The upper part of each panel overplots the normalized photometry for each visit (filled circles), with the first visit in black and the second visit in red, along with the best-fit transit light curves (solid lines) where the orbital inclination, $a/R_{\star}$, and transit time have been allowed to vary freely for each individual transit.  The lower panel takes the difference between the two light curves (black filled circles) and compares it to the difference between the best-fit transit solutions (solid black line).  Note that even when all transit parameters are allowed to vary freely, it is not possible to reproduce the sharp features visible during ingress and egress in the lower left panel.} 
\label{transit_diff}
\end{figure*}

One consequence of a misalignment between the star's rotation axis and the planet's orbit is that the planet will not necessarily occult the same spot on successive transits, as would be expected for a well-aligned system; we therefore consider each transit individually.  Our analysis above indicates that the 3.6~\micron~transit on UT 2009 Jan 28 displays a statistically significant increase in the standard deviation of the in-transit data that is dominated by contributions from correlated noise on time scales greater than 30~s, as would be expected if the planet occulted an active region on the surface of the star.  Although the 4.5~\micron~transit from UT 2009 Jan 30 does not appear to display a similar increase, our imperfect correction for the intrapixel sensitivity variations in this visit means that we are less sensitive to variations in $\sigma_\texttt{rel}$.  We argue that even if the star's rotation axis and the planet's orbit are misaligned, it is still likely that the planet would occult the same active region during both the UT 2009 Jan 28 and Jan 30 visits, as the interval between these visits is much shorter than the star's approximately 50 day rotation period.   As we discuss later in this section, the fact that both visits display increased transit depths and shifted transit times provides additional support for this hypothesis.

We also consider the possibility that the increased scatter in the in-transit residuals might be due to a change in the transit parameters, including the planet's radius, orbital inclination, transit time, or $a/R_{\star}$, from one visit to the next.  We test this hypothesis by taking the difference of the first and second visits in each bandpass from 2009 and comparing the shape of the residual light curve to the differences we would expect due to changes in these parameters, which should be distinct from the deviations created by occulted star spots (Fig. \ref{transit_diff}).  Because we are directly differencing the two light curves, our results are independent of any assumptions about the shape of the transit light curve or the stellar limb-darkening.  We inspect the deviations in the residuals plotted in Fig. \ref{transit_diff} and conclude that they do not appear to be well-matched by changes in the best-fit transit parameters, leaving occultations of active regions on surface of the star as the most likely hypothesis.  

If the planet occults a spot it can also cause a shift in the best-fit transit times, particularly when the spot is near the edge of the star and is occulted during ingress or egress.  Indeed, we see that the UT 2009 Jan 28 3.6~\micron~appears to occur $31.4\pm9.5$~s early, while the 4.5~\micron~Jan 30 visit occurs $34.4\pm9.4$~s late (see Fig. \ref{transit_o_c}) in the fits where we fix $a/R_{\star}$ and $i$ to a single common value.  As a test we repeated our fit to the 3.6~\micron~transit excluding the first 1/3 of the transit light curve, and found that the best-fit transit time shifted forward by approximately 30 s.  We would also expect that transits observed in visible light, where the contrast between the spots and the star is more pronounced, would show proportionally larger timing deviations when the planet crosses a spot.  As noted in \S\ref{timing}, the scatter in the measured visible-light transit times is inconsistent with a constant period, and the amplitude of the visible-light deviations is on average larger than the deviations in the infrared.  We should also see this same wavelength-dependence in the measured transit depths in Fig. \ref{depth_comparison}, and indeed we find that the 3.6~\micron~transit depth changes by 7.8\%, the 4.5~\micron~transit depth changes by 5.3\%, and the 8.0~\micron~transit depth changes by 4.9\% during the period between UT 2009 Jan 9 and Feb 2.  Lastly, we can examine the visible-light flux measurements for GJ~436 in Fig. \ref{stellar_rot} and see that these two transits were obtained near a minimum in the star's flux, consistent with a relative increase in the fractional spot coverage as compared to earlier epochs.  The measured values for \sval, a common activity indicator, appear to be anti-correlated with the observed flux variations and reach a local maximum near this point.

\subsection{Atmospheric Transmission Spectrum}\label{modeling_disc}

In principle, the broadband transmission photometry of GJ~436b allows us to constrain the chemical composition and  temperature structure near the limb of the planetary atmosphere \citep[e.g.,][]{madhu09}.  However, time variability in either the properties of the star or of the planet poses a significant challenge to an analysis in which we are comparing transit observations at different wavelengths obtained days or weeks apart.  As discussed in \S\ref{time_var_radius}, we consider it unlikely that the discrepancies in the measured transit depths are due to changes in the properties of the planet, but instead conclude in \S\ref{star_spots} that the occultations of regions of nonuniform brightness in a subset of the transits appear to be responsible for the observed depth variations.

\begin{figure*}
\epsscale{1.0}
\plotone{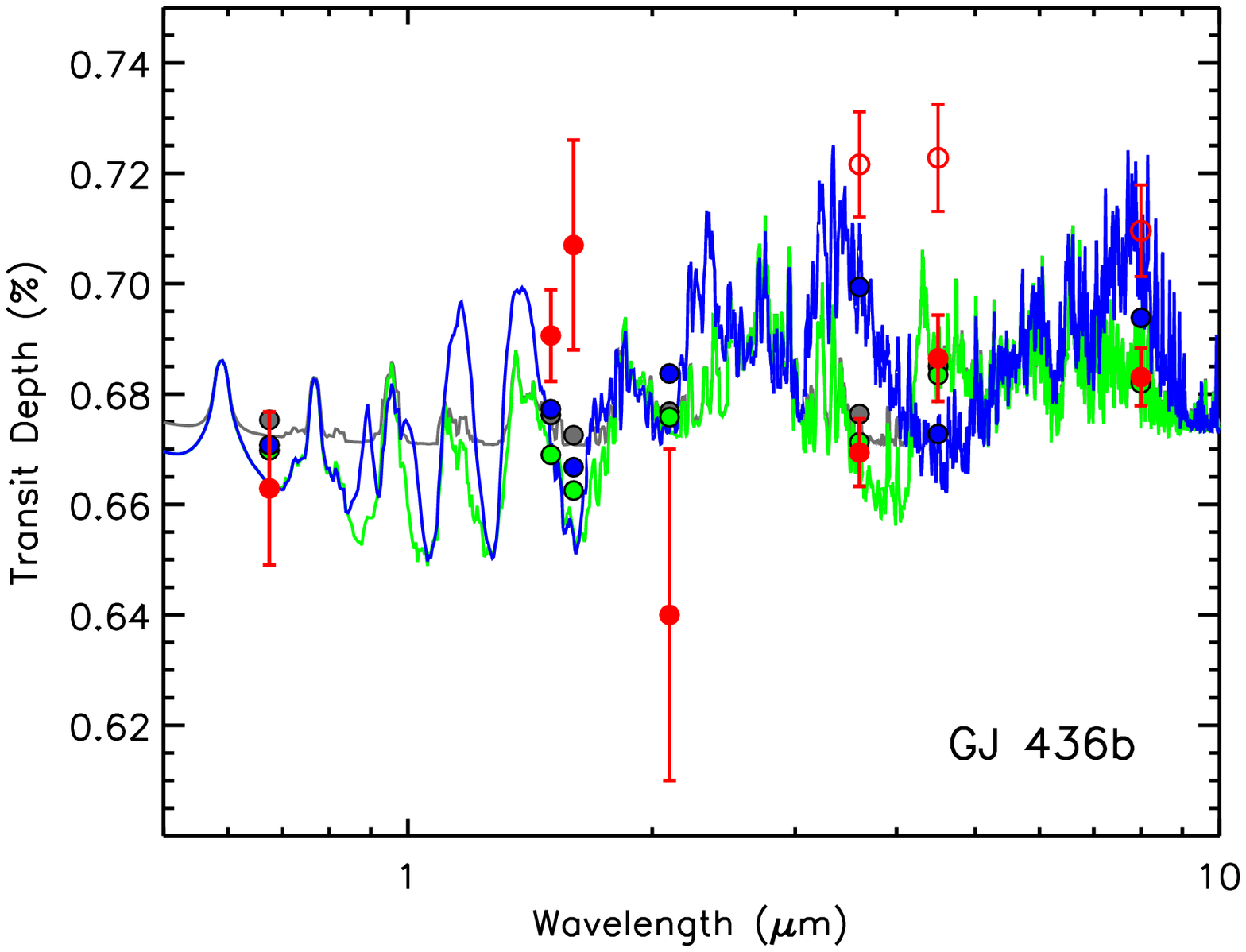}
\caption{Comparison between measured transit depths (red circles) and model transmission spectra, where the transit depth is defined as the square of the best-fit planet-star radius ratio in each band.  We include previously published visible and near-IR transit depths from (in order of increasing wavelength) \citet{ballard10}, \citet{pont08},  \citet{alonso08}, and \citet{caceres09} along with the \emph{Spitzer} transit depths at 3.6, 4.5, and 8.0~\micron.  Open circles indicate \emph{Spitzer} observations in which the planet appears to transit regions of non-uniform brightness on the star, as discussed in \S\ref{star_spots}.  Model transmission spectra include an atmosphere with reduced methane and enhanced CO abundances \citep{stevenson10} in green, a methane-rich model similar to that described in \citet{beaulieu10} in blue, and a methane-poor model with an opaque cloud deck at pressures below 50 mbar in grey, which provides a better match to the visible-light transit depth from \citet{ballard10}.  All models are calculated using the methods described in \citet{madhu09}, where we fix the dayside $P$-$T$ profiles to the nominal best-fit profile from \citet{stevenson10}.  We find that allowing the $P$-$T$ profile to vary freely in our fits has a negligible effect on the agreement between the data and the green best-fit model.  Colored green, blue, and grey circles indicate the predicted values for these models integrated over the bandpasses of the observations.}
\label{atm_models}
\end{figure*}

If we set aside those transits which we believe to be most strongly affected by stellar activity, including the UT 2009 Jan 28 and 30 visits, we may attempt to estimate the shape of the planet's transmission spectrum using the remaining transits.  Although the evidence for spots in the final 8.0~\micron~transit on UT 2009 Feb 2 is somewhat weaker, we choose to exclude it on the grounds that it displays some of the same behaviors (increased depth, larger than usual timing offset) as the more strongly affected 3.6 and 4.5~\micron~transits immediately preceding it.  If we then average the remaining three 8.0~\micron~depths, we find depths of [$0.6694\%\pm0.0061\%$,$0.6865\%\pm0.0078\%$,$0.6831\%\pm0.0052\%$] at 3.6, 4.5, and 8.0~\micron, respectively.  These three values are consistent with the near-IR transit depth from \citet{pont08} of $0.6906\%\pm0.0083\%$ ($1.1-1.9$~\micron), as well as the best-fit visible light transit depth from \citet{ballard10}, $0.663\%\pm0.014\%$ ($0.35-1.0$~\micron).  Ground-based data provide additional constraints in the near-IR, including a $H$ band transit depth of $0.707\%\pm0.019\%$ from \citet{alonso08} and a $Ks$ transit depth of $0.64\%\pm0.03\%$ from \citet{caceres09}\footnote[1]{The best-fit planet-star radius ratio reported by these authors is inconsistent with their best-fit depth.  We re-fit their data with an equivalent model and conclude that this discrepancy is most likely the result of a mistake in the reported value for the radius ratio, as our best-fit depth is a good match for the value stated in the paper.}, both from individual transit observations.

We fit these data using the retrieval technique described in \citet{madhu09}, which explores the parameter space of a one-dimensional, hydrogen-rich model atmosphere. We compute line-by-line radiative transfer with the assumption of hydrostatic equilibrium and we use parametric prescriptions for the relative abundances of H$_2$O, CH$_4$, CO, CO$_2$. We also include other dominant visible-light and infrared opacity sources, including Na, K, H$_2$-H$_2$ collision-induced absorption, and Rayleigh scattering. Our molecular line data are from \citet{rothman05}, \citet{freedman08}, Freedman (personal communication, 2009), \citet{karkoschka10}, and Karkoschka (personal communication, 2011). The H$_2$-H$_2$ opacities are from \citet{borysow97} and \citet{borysow02}. We fix the pressure-temperature ($P$-$T$) profile to the best-fit dayside profile from \citet{stevenson10} and \citet{madhu10}; it is possible to obtain a marginally improved fit to these data if we allow the $P$-$T$ profile to vary freely in the fit, but the differences are not significant. We find that the observations can be explained to within the 1-$\sigma$ uncertainties by a methane-poor model (green line in Fig. \ref{atm_models}) that contains mixing ratios of H$_2$O = $1.0\times 10^{-3}$, CO = $1.0\times 10^{-3}$, and CH$_4$ = $1.0\times 10^{-6}$; the data used in this fit appear to be inconsistent with methane abundances $\geq 10^{-5}$. This model also includes CO$_2$ = $1.0\times 10^{-5}$, but the concentration of this molecule is less well constrained, as it is degenerate with the CO abundance in the 4.5~\micron~band.  We do not expect strong absorption due to atomic Na and K in this temperature regime \citep{sharp07}, and we therefore adopt Na and K mixing ratios of $0.1 \times $ solar abundances.  If we compare the visible-light transit depth of 0.650\% from this model to the value reported by \citet{ballard10} we find that it is consistent at the $0.5\sigma$ level.  Model transmission spectra for GJ~436b from \citet{shabram10}, such as the rescaled model including higher-order hydrocarbons (model ``g" in Shabram et al.) also provide a reasonably good match to these data.

We can reduce the disagreement between the measured transit depths and the green model  in the $1-2$~\micron~wavelength range by introducing an opaque cloud layer at 50 mbar (grey model in Fig. \ref{atm_models}).  However, such a cloud layer would be inconsistent with the dayside emission spectrum measured by \citet{stevenson10} unless it was optically thin in the center of the dayside hemisphere, or only intermittently present as discussed in \S\ref{time_var_radius}. We also note that occultations of spots and other features on the star will have a stronger effect on the measured transit depth at shorter wavelengths, and it is therefore possible that these measurements (several of which were derived from individual transit observations) are unreliable for our purposes here.

Returning to the \emph{Spitzer} data, we find that our conclusions about the atmospheric composition are strongly dependent on our choice of which transit depths to include in our analysis.  We illustrate this with a blue model in Fig. \ref{atm_models}, which contains H$_2$O and CH$_4$ mixing ratios of $5.0\times 10^{-4}$ each and no CO or CO$_2$, and is comparable to the model presented in \citet{beaulieu10}.  \citet{beaulieu10} excluded the shallower 3.6~\micron~transit on UT 2009 Jan 9 and kept the deeper 3.6~\micron~UT 2009 Jan 28 and 8.0~\micron~UT 2009 Feb 2 visits in their analysis, and as a result they concluded that the planet's transmission spectrum contained strong methane features, as illustrated by this blue model.  They argue that the correction for the intrapixel effect is degenerate with the transit depth for the UT 2009 Jan 9 visit and that this visit is therefore unreliable, but we find that there is good overlap between the $x$ and $y$ positions spanned by the in-transit and out-of-transit data.  We obtain transit depths that are consistent at the $0.1\sigma$ level when we fit for our intrapixel sensitivity correction using either the entire light curve or the out-of-transit data alone.  Although our 3.6 and 8.0~\micron~transit depths are in good agreement with the values obtained by Beaulieu et al., our best-fit transit depth for the 4.5~\micron~UT 2009 Jan 17 is $2.5\sigma$ larger.  We note that Beaulieu et al. allow $a/R_{\star}$ and $b$ to vary individually for each transit, and that their values for these parameters from the Jan 17 transit fit are outliers when compared to other visits; we conclude that this is likely the cause of their shallower best-fit radius ratio.  Despite this disagreement, we find that if we include the same transits as Beaulieu et al. in our analysis, we also produce a transmission spectrum that is consistent with strong methane absorption. 

\begin{figure}
\epsscale{1.2}
\plotone{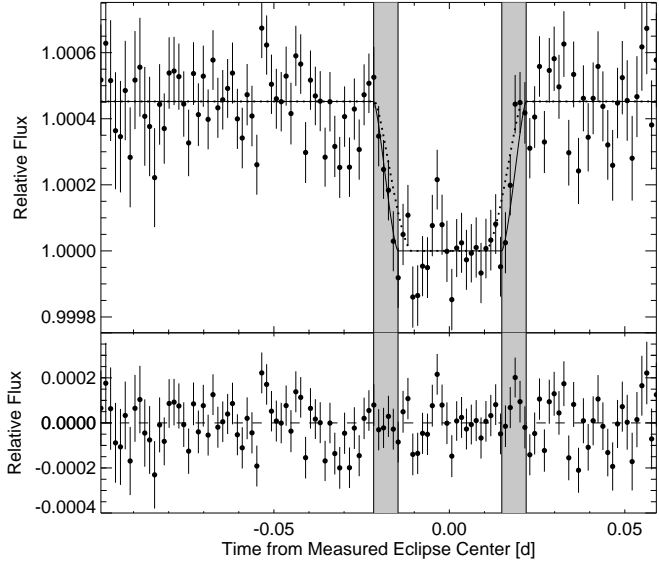}
\caption{Photometry for eleven 8~\micron~secondary eclipses (filled circles), with detector effects removed.  Individual visits have been aligned using the best-fit transit ephemeris and assuming a constant offset for the secondary eclipse.  The best-fit secondary eclipse light curve is overplotted (solid line), and residuals from this curve are shown in the lower panel.  The period spanned by ingress and egress is denoted as a grey shaded region; we find no significant deviations from a model in which the planet has a uniform surface brightness.  The dotted line shows the best-fit transit light curve, rescaled to match the depth of the secondary eclipse shown here.  The longer ingress and egress for the transit are due to the increased planet-star distance and correspondingly higher impact parameter during this event, which occurs close to apastron.}
\label{combined_eclipse}
\end{figure}

\begin{figure}
\epsscale{1.2}
\plotone{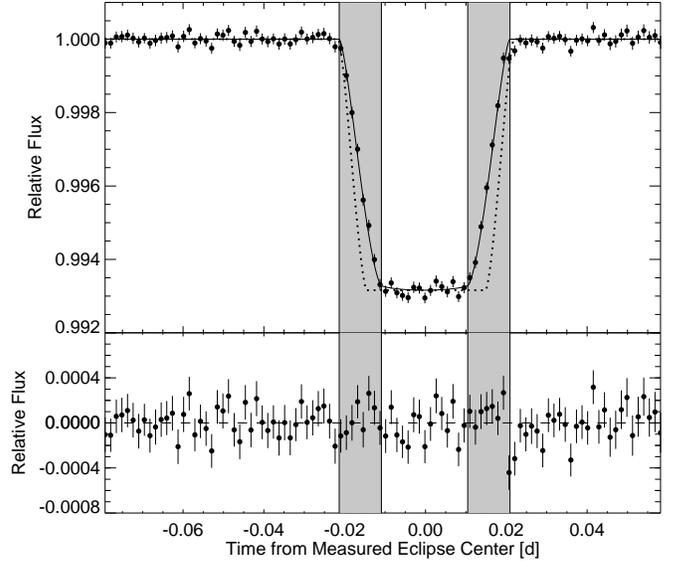}
\caption{Photometry for four 8~\micron~transits (filled circles), with detector effects removed and light curves aligned using the best-fit transit ephemeris.  The best-fit transit light curve is overplotted (solid line), and residuals from this curve are shown in the lower panel.  The period spanned by ingress and egress is denoted as a grey shaded region; we find no significant deviations from the expected spherical planet model, as might be expected if the planet was significantly oblate.  The dotted line shows the best-fit secondary eclipse light curve, rescaled to match the depth of the transit light curve where the limb-darkening is set to zero.  The shorter ingress and egress for the secondary eclipse are due to the reduced planet-star distance and correspondingly lower impact parameter during this event, which occurs shortly before periastron.}
\label{combined_transit}
\end{figure}

If, as we propose, occulted regions of non-uniform brightness on the surface of the star are responsible for the discrepancies in the 3.6 and 4.5~\micron~transit depths, it will be difficult to provide a definitive characterization of GJ~436b's transmission spectrum with broadband \emph{Spitzer} photometry.  Our analysis suggests that the atmosphere of GJ~436b is likely under-abundant in methane and over-abundant in CO, consistent with the conclusions of \citet{stevenson10} and \citet{madhu10}, but in order to reach these conclusions we have assumed that we have correctly identified and excluded all transits in which the planet occults active regions on the star.  However, if the fractional spot coverage on the star is sufficiently high, it is possible that \emph{all} transits are affected by these regions, in which case we cannot draw any robust conclusions about the shape of the planet's transmission spectrum.

\subsection{Dayside Emission Spectrum and Limits on Variability}\label{dayside_var}

We can use the eleven secondary eclipse depths listed in Table \ref{eclipse_param} to study the properties of the planet's dayside atmosphere.  We take the error-weighted average of the eclipse depths and find a combined value of $0.0452\%\pm0.0027\%$, consistent with the value of $0.054\%\pm0.008\%$ reported by \citet{stevenson10}.  Next we construct a combined light curve incorporating all eleven secondary eclipse observations, shown in Fig. \ref{combined_eclipse}.  Fig. \ref{combined_transit} shows the equivalent combined 8~\micron~transit light curve for comparison.  As a check we fit these combined data with a secondary eclipse light curve and find that the best-fit eclipse depth agrees exactly with this error-weighted average from the individual eclipse fits.  Because the strongest constraints on the relative abundances of methane and CO come from the 3.6 and 4.5~\micron~eclipse measurements, we do not expect the reduced 8~\micron~error bar to affect the conclusions reached by Stevenson et al. regarding these molecules.  If we compare our results to the two models plotted in Fig. 2 of Stevenson et al., we find that the revised 8~\micron~eclipse depth is best-described by a cooler model with an effective blackbody temperature of 790~K (defined as the temperature needed to match the total integrated flux at all wavelengths) and a modestly enhanced (30x higher) water abundance, rather than the hotter 860~K model with weaker water absorption.  We also calculate a revised brightness temperature for the planet in the 8~\micron~band, defined as the temperature required to match the observed planet-star flux ratio in this bandpass assuming that the planet radiates as a blackbody.  We use the parameters in Table \ref{global_param} and assume a \texttt{Phoenix} atmosphere model with an effective temperature of 3585~K and $\log(g)$ equal to 4.843 \citep{torres07} for the star, and find that the planet has a best-fit brightness temperature of $740\pm16$~K.

Returning to Fig. \ref{combined_eclipse}, we examine the residuals from our best-fit eclipse solution to search for evidence of deviations during ingress and egress caused by a non-uniform day-side surface brightness \citep{williams06,rauscher07}.  The primary effect of a non-uniform brightness distribution is to shift the best-fit eclipse time \citep[e.g.,][]{agol10}, but in this case uncertainties in estimates for GJ~436b's orbital eccentricity and longitude of periastron prevent us from detecting the small ($<1$ minute) timing offsets expected from this effect.  This timing offset will also display a small wavelength-dependence, due to variations in the brightness distribution as seen in different bandpasses, but this signal is likely to be too weak to detect by comparing to the existing 3.6 and 5.8~\micron~eclipse observations from \citet{stevenson10}.  Instead, we seek to determine if the shape of the 8~\micron~eclipse ingress and egress can be used to constrain the planet's day-side brightness distribution.  We compare the eclipse light curves for a uniform surface brightness disk to that of a local equilibirum model \citep[i.e., one with the radiative time set to zero so that each region of the planet is at its local equilibrium temperature;][]{hansen08,burrows08}, and find that the peak-to-trough residuals between these light curves is only $0.002\%$, if the eclipse depth is a free parameter.  This is approximately a factor of ten smaller than our measurement errors, as demonstrated by the binned residuals in Fig. \ref{combined_eclipse}.  As we increase the amount of energy advected to the planet's night side using the models described in \citet{cowan10}, the location of the hot spot on the planet's day side shifts away from the substellar point and the overall temperature contrast decreases.  Because we are not sensitive to the timing offset caused by the shifted hot spot, the only effect of this increased advection is to homogenize the planet's temperatures, producing light curves increasingly similar to the uniform disk light curves.  

\begin{figure}
\epsscale{1.2}
\plotone{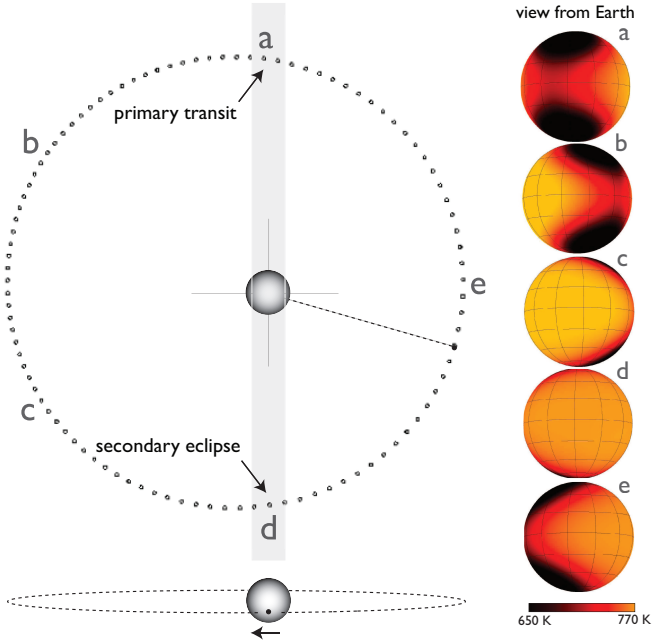}
\caption{The left panel shows an orbital diagram for the GJ 436 system.  Distances and radii are drawn to scale, and the location of periastron is marked by a dotted line.  Grey shaded regions indicate the locations of the planet during the transit and secondary eclipse, where the viewer is assumed to be at the top of the plot.  The right-hand panel shows five snapshots from a general circulation model for this planet as seen at different orbital phases by an observer on the Earth.}
\label{orbit_diagram}
\end{figure}

\subsubsection{A Variability Study for GJ~436b}

Tidal dissipation is expected to have driven GJ~436b into a pseudo-synchronous rotation state in which the planet's spin frequency is nearly commensurate with the planet's instantaneous orbital frequency at periastron. There are several competing theories of the pseudosynchronization process \citep[see, e.g.][]{ivanov07}. We adopt the expression given by \citet{hut81}:
\begin{align}\label{eq6}
{\Omega_{\rm spin}\over{\Omega_{\rm orbit}}}={
1+{15\over{2}}e^{2}+{45\over{8}}e^{4}+{5\over{16}}e^{6}
\over{(1+3e^2+{3\over{8}}e^{4})(1-e^{2})^{3/2}}}\, .
\end{align}
For GJ~436b, this relation gives $P_{\rm spin}=2.32$~days, which yields a 19-day synodic period
for the star as viewed from a fixed longitude on the planet. GJ~436b also experiences an 83\% increase in incident flux during the 1.3-day interval between apoastron and periastron.

\begin{figure}
\epsscale{1.1}
\plotone{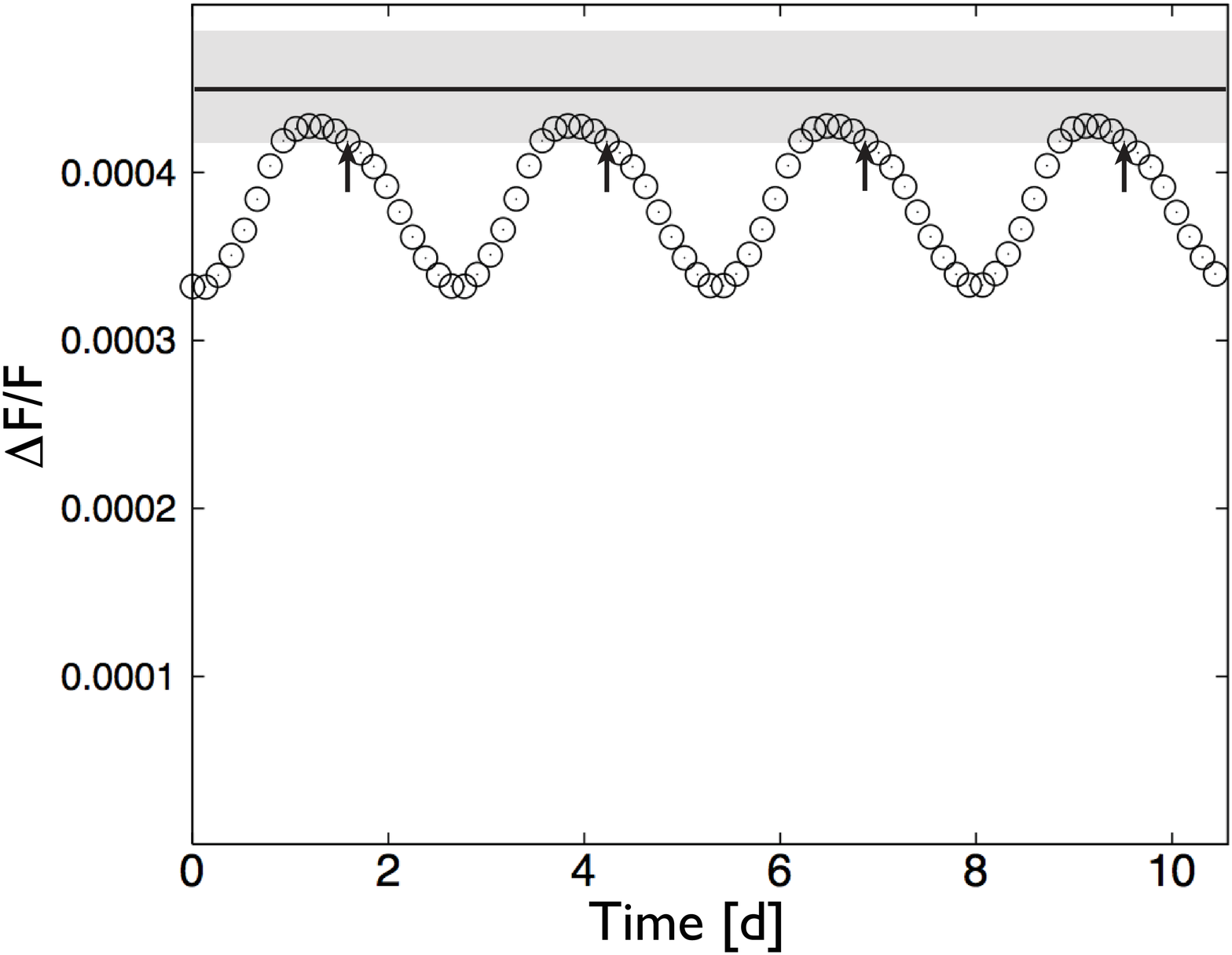}
\caption{Predicted 8~\micron~emission for GJ~436b (open circles) as a function of time, from general circulation models described in \citet{langton08}.  The periodic modulation in flux is primarily due to the changing orbital geometry as we watch the hotter dayside rotate in and out of view, with a secondary effect caused by the heating and cooling of the atmosphere as the planet moves from periastron to apastron and back.  The predicted fluxes during the secondary eclipse, as indicated by the black arrows, are nearly constant in time.  The horizontal black line and grey shaded region indicate the average secondary eclipse depth and corresponding $1\sigma$ uncertainty.}
\label{circ_model}
\end{figure}

We have computed simple hydrodynamical models to assess whether the asynchronous rotation and time-varying insolation  are likely to generate atmospheric flows that are sufficiently chaotic to produce observable orbit-to-orbit variability in the secondary eclipse depths.  Our two-dimensional hydrodynamical model contains three free parameters. The first, $p_{8 \mu \rm {m}}$, is the atmospheric pressure at the 8~\micron~photosphere; the second, $X$, corresponds to the fraction of the incoming optical flux that is absorbed at or above the 8~\micron~photosphere; and the third, $p_b$, corresponds to the pressure at the base of our modeled layer.  We adopt parameter values of $p_{8 \mu \rm {m}}$ = 100 mbar, $p_b$ = 4.0 bar and $X$ = 1.0 for these models, which puts our modelÕs light curve in good agreement with GJ~436b's average 8~\micron~secondary eclipse depth. The full details of the computational scheme are the same as those adopted in \citet{langton08}, with updates as described in \citet{laughlin09}.  A model photometric light curve is then obtained by integrating at each time step over the planetary hemisphere visible from Earth, where we assume that each patch of the planet radiates with a black-body spectrum corresponding to the local temperature.

The model is run for a large number of orbits, and a quasi-steady state surface flow emerges. The temperature structure of this flow as seen from an observer in the direction of Earth at five equally spaced intervals in the orbit is shown in Figure \ref{orbit_diagram}, and the model light curve over these five orbits is shown in Figure \ref{circ_model}. Over the course of a single orbit, the $8~\micron$ planet-to-star flux ratio varies nearly sinusoidally from $\Delta F/F=0.033\%$ to $0.043\%$. The model's flux at secondary eclipse agrees well with the observed value, and varies by only 0.5\% peak-to-peak from one orbit to the next.  We note that more sophisticated three-dimensional general circulation models for GJ~436b from \citet{lewis10} also predict very low ($1.3-1.5\%$) levels of variability in the 8~\micron~band for a range of atmospheric metallicities. 

Although these models indicate that GJ~436b's modest orbital eccentricity is likely not sufficient to induce significant variability, they also do not include many processes such as clouds, photochemistry, and small scale turbulence that are known to contribute to temporal variability in planetary atmospheres.  We therefore place empirical limits on GJ 436b's dayside variability using the eleven 8~\micron~secondary eclipse observations.  We assume that the intrinsic dayside fluxes are drawn from either a Gaussian distribution with a standard deviation $\delta$, or from a boxcar distribution with a width equal to $2\delta$.  In both cases we set the mean of the distribution equal to the error-weighted mean of the measured secondary eclipse depths given in Table \ref{global_param}.  We then conduct 10,000 random trials, where we draw eleven measurements from each distribution and calculate the reduced $\chi^2$ of these values as compared to the measured secondary eclipse depths in Table \ref{eclipse_param}.  We then determine the fraction of the 10,000 random trials in which the reduced $\chi^2$ is less than or equal to one, which should correspond to the probability that the underlying distribution is consistent with the measured eclipse depths.  We repeat this calculation for a range of values for $\delta$, and plot the resulting probability distribution as a function of $\delta$ for both boxcar and Gaussian distributions.  We find that for a boxcar distribution we can place [$1\sigma$, $2\sigma$, $3\sigma$] limits on the intrinsic variability of [29\%, 42\%, 58\%], and for a Gaussian distribution our corresponding upper limits are [17\%, 27\%, 42\%].  These limits are consistent with the predictions from general circulation models for this planet, but they are not low enough to provide meaningful constraints on these models.
 
\section{Conclusions}

In this paper we present \emph{Spitzer} observations of eight transits and eleven secondary eclipses of GJ~436b at 3.6, 4.5, and 8.0~\micron, which allow us to derive improved values for the planet's orbital ephemeris, eccentricity, inclination, radius, and other system parameters.  We discuss the effects that our assumptions about the longitude of periastron and stellar limb-darkening profiles have on our best-fit transit parameters, and find that our best-fit parameters vary by $1\sigma$ or less in all cases.  We find that all parameters are consistent with a constant value over the two-year period spanned by our observations, with the exception of the measured transit depths and times in the 3.6 and 4.5~\micron~bands.  We find that the 3.6~\micron~radius ratio measured on UT 2009 Jan 28 is $4.7\sigma$ deeper than the value measured on UT 2009 Jan 9 in this same band, and the 4.5~\micron~radius ratio from UT 2009 Jan 30 is $2.9\sigma$ deeper than the value measured on UT 2009 Jan 17.  The level of significance for these changing radius ratios remains high even after accounting for the effects of residual correlated noise in the data.  

We also present an improved estimate for GJ~436b's 8~\micron~secondary eclipse depth, based on eleven eclipse observations in this bandpass.  We find that the new depth is consistent with previous models described in \citet{stevenson10} and \citet{madhu10}, although we prefer solutions with modestly lower effective temperatures (790~K instead of 860~K).  We use the shape of the eclipse ingress and egress to search for the presence of a non-uniform temperature distribution in the planet's dayside atmosphere, but uncertainties in the predicted time of secondary eclipse ultimately limits our ability to place meaningful constraints on this quantity.  Our eclipse depths in this band are consistent with a constant value, and we place a $1\sigma$ upper limit of 17\% on variability in the planet's dayside atmosphere.  This limit is in good agreement with the predictions of general circulation models for this planet, which are typically variable at the level of a few percent or less in this bandpass. 

Although it is possible that such residual noise or a time-varying cloud layer at sub-mbar pressures could explain the apparent transit depth variations, the features observed in the transit light curves appear to be most consistent with the presence of occulted spots or other areas of non-uniform brightness on the surface of the star in the UT 2009 Jan 28 and 30 transits.  We find that for the UT 2009 Jan 28 transit the in-transit data have a higher RMS than the out-of-transit data, as would be expected for occulted spots; we would expect poorly corrected systematics to produce an equivalently large RMS in both the in-transit and out-of-transit data, as the star spans same region of the pixel in both segments.  Although we are not as sensitive to such effects in the UT 2009 Jan 30 visit, which has higher levels of correlated noise due to an imperfect correction for intrapixel sensitivity variations, the short separation between these two observations relative to the star's approximately 50 day rotation period means that the planet is likely to have occulted the same feature in both visits.  We also see statistically significant variations in the measured transit times, where the amplitude of the variations is typically smaller for infrared observations than for those obtained in visible light, also suggesting the present of occulted spots.  We note that the anomalously deep transits observed on UT 2009 Jan 28 and 30 also have best-fit transit times that are offset by 30 s ($3.1-3.5\sigma$ significance) from the predicted values.  The fact that the three deepest transits are all measured within the same five-day period is also consistent with a single epoch of increased stellar activity.  We reconcile this conclusion with the absence of any variations larger than a few mmag in the star's visible and infrared fluxes by proposing that the star's spin axis is likely inclined with respect to our line of sight, which has the effect of reducing the amplitude of any flux variations independent of spot coverage.  If this is in fact the case, GJ~436b's orbit will be misaligned with respect to the star's spin axis.   

If we examine the wavelength-dependent transit depths for the subset of visits that appear to be least affected by spots, we find that the resulting transmission spectrum is consistent with the same reduced methane and enhanced CO abundances used by \citet{stevenson10} to fit the planet's dayside emission spectrum.  These same transit data are also consistent with models including an opaque cloud layer at a pressure of approximately 50 mbar or less in the planet's atmosphere, which reduces the amplitude of the absorption features in the model spectra.  We find no convincing evidence for the strong methane absorption reported by \citet{beaulieu10}, although we note that our conclusions vary significantly depending on which transits we include in our analysis.  It is possible that all measured transit depths are affected to varying degrees by stellar activity, in which case it may not be feasible to characterize the planet's transmission spectrum using broadband photometry obtained over multiple epochs.  Because active regions occulted by the planet display a characteristic wavelength-dependence and also alter the local shape of the transit light curve, high signal-to-noise grism spectroscopy of the transit over multiple epochs would help to resolve this issue.  Such observations would also provide an independent test of the reliability of the \emph{Spitzer} transit data; if similar apparent depth variations were observed in other data sets, it would provide a strong argument against the hypothesis that the apparent depth variations in these data might be the result of poorly corrected instrument effects.  Lastly, grism spectroscopy could also be used to search for time-varying clouds at sub-mbar pressures, which should produce a featureless transmission spectrum with a uniformly increased depth when present, as compared to the standard cloud-free transmission spectrum.

As indicated by its rotation rate and \caii~emission, GJ~436 is an old and relatively quiet early M star.  If the apparent transit depth variations we describe here are indeed due to the occultation of active regions on the star, as appears likely, we would expect similar features to occur frequently in the transit light curves of other planets orbiting M dwarfs at all activity levels.  GJ 1214 is currently the only other M star known to host a transiting planet, and has a similar 53-day rotation period and a modestly lower 3000~K effective temperature as compared to GJ~436 \citep{charbonneau09,berta10}.  A majority of the published data on this system are in the visible and near-infrared wavelengths where star spots should be prominent, and several recent papers report the presence of occulted spots in a subset of transit observations \citep{berta10,carter10,kundurthy10}.  Such spots might also account for the apparent disagreement in measurements of the planet's infrared transmission spectrum, which some authors find to be featureless \citep{bean10,desert11}, while others detect absorption features \citep{croll11}.  HD~189733b is currently the only other exoplanet with repeated \emph{Spitzer} transit observations in the same band; although this planet orbits a relatively active K star \citep[e.g.,][]{knutson10}, it exhibits much smaller variations in the measured transit depths and times as compared to GJ~436b \citep{agol10,desert10}.  This is perhaps not surprising, as the relative fractional spot coverage, spot sizes, and spot temperatures may well be qualitatively different on K stars and M stars.

\acknowledgments

We would like to thank the anonymous referee for a very thoughtful report, as well as Jonathan Fortney, Megan Shabram, and Nikole Lewis for helpful discussions on the implications of our data for their published models of GJ~436b.  We are also grateful to Eric Gaidos for his commentary on the nature of activity on M dwarfs, and Josh Winn for helpful discussions on spin-orbit alignment for GJ~436b.  We would also like to thank Howard Isaacson for supplying the \sval~values for our activity study, and to acknowledge the Keck observers who obtained the HIRES spectra used for these measurements, including Andrew Howard, John Johnson, Debra Fischer, and Geoff Marcy.  This work is based on observations made with the \emph{Spitzer Space Telescope}, which is operated by the Jet Propulsion Laboratory, California Institute of Technology, under a contract with NASA.  Support for this work was provided by NASA through an award issued by JPL/Caltech.  HAK was supported by a fellowship from the Miller Institute for Basic Research in Science.  EA was supported in part by the National Science Foundation under CAREER Grant No. 0645416.


\begin{thebibliography}{}

\bibitem[Adams et al.(2008)]{adams08} Adams, E., Seager, S., \& Elkins-Tanton, L. 2008, \apj, 673, 1160
\bibitem[Agol et al.(2010)]{agol10} Agol, E., et al. 2010, \apj, 721, 1861
\bibitem[Alonso et al.(2008)]{alonso08} Alonso, R., et al. 2008, \aap, 487, L5
\bibitem[Bakos et al.(2006)]{bakos06} Bakos, G. \'A , et al. 2006, \apj, 650, 1160
\bibitem[Ballard et al.(2010a)]{ballard10a} Ballard, S., et al. 2010a, \apj, 716, 1047 
\bibitem[Ballard et al.(2010b)]{ballard10} Ballard, S., et al. 2010b, \pasp, 122, 1341
\bibitem[Basri et al.(2011)]{basri10} Basri, G., et al. 2011, \apj, 141, 20
\bibitem[Batygin et al.(2009)]{batygin09} Batygin, K., et al. 2009, \apj, 699, 23
\bibitem[Bean \& Seifahrt(2008)]{bean08} Bean, J. L. \& Seifahrt, A. 2008, \aap, 487, L25
\bibitem[Bean et al.(2010)]{bean10} Bean, J. L., Kempton, E. M.-R., \& Homeier, D. Nature, 468, 669
\bibitem[Beaulieu et al.(2011)]{beaulieu10} Beaulieu, J.-P., et al. 2011, \apj, 731, 16
\bibitem[Berta et al.(2011)]{berta10} Berta, Z. K., et al. 2011, \apj submitted, arXiv:1012.0518
\bibitem[Borysow et al.(1997)]{borysow97} Borysow, A., Jorgensen, U. G., \& Zheng, C. 1997, A\&A, 324, 185.
\bibitem[Borysow(2002)]{borysow02} Borysow, A. 2002, A\&A, 390, 779
\bibitem[Burrows et al.(2008)]{burrows08} Burrows, A., Budaj, J., \& Hubeny, I. 2008, \apj, 678, 1436
\bibitem[Butler et al.(2004)]{butler04} Butler, R. P., et al. 2004, \apj, 617, 580
\bibitem[C\'aceres et al.(2009)]{caceres09} C\'aceres, C., et al. 2009, \aap, 507, 481
\bibitem[Carter et al.(2011)]{carter10} Carter, J. A., et al. 2011, \apj, 730, 82
\bibitem[Charbonneau et al.(2002)]{charbonneau02} Charbonneau, D., et al. 2002, \apj, 568, 377
\bibitem[Charbonneau et al.(2005)]{charbonneau05} Charbonneau, D., et al. 2005, \apj, 626, 523
\bibitem[Charbonneau et al.(2008)]{charbonneau08} Charbonneau, D. et al. 2008, \apj, 686, 1436
\bibitem[Charbonneau et al.(2009)]{charbonneau09} Charbonneau, D., et al. 2009, \nat, 462, 891
\bibitem[Claret(2000)]{claret00} Claret, A. 2000, \aap, 363, 1081
\bibitem[Claret(2008)]{claret08} Claret, A. 2008, \aap, 482, 259
\bibitem[Claret(2009)]{claret09} Claret, A. 2009, \aap, 506, 1335
\bibitem[Claret \& Hauschildt(2003)]{claret03} Claret, A., \& Hauschildt, P. H. 2003, \aap, 412, 241
\bibitem[Coughlin et al.(2008)]{coughlin08} Coughlin, J. L., et al. 2008, \apj, 689, L149
\bibitem[Cowan, Agol, \& Charbonneau(2007)]{cowan07} Cowan, N. B., Agol, E., \& Charbonneau, D. 2007, \mnras, 379, 641
\bibitem[Cowan \& Agol(2011)]{cowan10} Cowan, N. B. \& Agol, E. 2011, \apj, 726, 82
\bibitem[Croll et al.(2011)]{croll11} Croll, B., et al. 2011, \apj in press, arXiv:1104.1011
\bibitem[Crossfield et al.(2010)]{crossfield10} Crossfield, I. J. M., et al. 2010, \apj, 723, 1436
\bibitem[Cushing et al.(2005)]{cushing05} Cushing, M. C., Rayner, J. T., \& Vacca, W. D. \apj, 623, 1115
\bibitem[de Kort(1954)]{dekort54} de Kort, J. J. M. A. 1954, Ricerche astronomiche, vol. 3, n. 5, Citta del Vaticano : Specola vaticana, 1954., p. 109
\bibitem[Deming et al.(2005)]{deming05} Deming, D., Seager, S., \& Richardson, L. J. 2005, \nat, 434, 740
\bibitem[Deming et al.(2006)]{deming06} Deming, D., Harrington, J., Seager, S., \& Richardson, L. J. 2006, \apj, 644, 560
\bibitem[Deming et al.(2007)]{deming07} Deming, D., et al. 2007, \apj, 667, L199
\bibitem[Deming et al.(2010)]{deming11} Deming, D., et al. 2011, \apj, 726, 95
\bibitem[Demory et al.(2007)]{demory07} Demory, B.-O., et al. 2007, \aap, 475, 1125
\bibitem[D\'esert et al.(2008)]{desert08} Desert, J.-M., et al. 2008, \aap, 492, 585
\bibitem[D\'esert et al.(2009)]{desert09} Desert, J.-M., et al. 2009, \apj, 699, 478
\bibitem[D\'esert et al.(2011a)]{desert10} Desert, J.-M., et al. 2011, \aap, 526, A12
\bibitem[D\'esert et al.(2011b)]{desert11} Desert, J.-M., et al. 2011, \apj, 731, L40
\bibitem[Eastman et al.(2010)]{eastman10} Eastman, J., Siverd, R., \& Gaudi, B. S. 2010, \pasp, 122, 894
\bibitem[Eaton et al.(2003)]{eat03} Eaton, J. A., Henry, G. W., \& Fekel, F. C. 2003, in The Future of Small Telescopes in the New Millennium, Volume II - The Telescopes We Use, ed. T. D. Oswalt (Dordrecht: Kluwer), 189
\bibitem[Fazio et al.(2004)]{faz04} Fazio, G.~G., et al., 2004, \apjs, 154, 10
\bibitem[Figueira et al.(2009)]{figueira09} Figueira, P. et al. 2009, \aap, 493, 671
\bibitem[Ford(2005)]{ford05} Ford, E. 2005, \apj, 129, 1706
\bibitem[Freedman et al.(2008)] {freedman08} Freedman, R. S., Marley, M. S. \& Lodders, K. 2008, \apj S, 174, 504
\bibitem[Fressin et al.(2010)]{fressin10} Fressin, F., et al. 2010, \apj, 711, 374
\bibitem[Gibson et al.(2011)]{gibson10} Gibson, N. P., Pont, F., \& Aigrain, S. 2011, \mnras, 411, 2199 
\bibitem[Gillon et al.(2007a)]{gillon07a} Gillon, M., et al. 2007a, \aap, 471, L51
\bibitem[Gillon et al.(2007b)]{gillon07b} Gillon, M., et al. 2007b, \aap, 472, L13
\bibitem[Grillmair et al.(2008)]{grillmair08} Grillmair, C. J. et al. 2008, \nat, 456, 767
\bibitem[Guillot \& Havel(2011)]{guillot10} Guillot, T. \& Havel, M. 2010, \aap, 527, A20
\bibitem[Hansen(2008)]{hansen08} Hansen, B. M. S. 2008, \apj, 179, 484
\bibitem[Harrington et al.(2007)]{harrington07} Harrington, J., et al. 2007, \nat, 447, 691
\bibitem[Hauschildt et al.(1999)]{hauschildt99} Hauschildt, P., Allard, F., Ferguson, J., Baron, E., \& Alexander, D. R. 1999, \apj, 525, 871
\bibitem[Henry(1999)]{henry99} Henry, G. W. 1999, PASP, 111, 845
\bibitem[Henry \& Winn(2008)]{henry08} Henry, G. W. \& Winn, J. N. 2008, \aj, 135, 68
\bibitem[Hut(1981)]{hut81} Hut, P.\ 1981, \aap, 99, 126 
\bibitem[Iro \& Deming(2010)]{iro10} Iro, N., \& Deming, L. D., 2010, \apj, 712, 218
\bibitem[Isaacson \& Fischer(2010)]{isaacson10} Isaacson, H., \& Fischer, D. 2010, \apj, 725, 875
\bibitem[Ivanov \& Papaloizou(2007)]{ivanov07} Ivanov, P.~B., \& Papaloizou, J.~C.~B.\ 2007, \mnras, 376, 682 
\bibitem[Jenkins et al.(2009)]{jenkins09} Jenkins, J. S., et al. 2009, \apj, 704, 975
\bibitem[Karkoschka \& Tomasko(2010)]{karkoschka10} Karkoschka, E. \& Tomasko, M. 2010, Icarus, 205, 674
\bibitem[Knutson et al.(2007)]{knutson07} Knutson, H. A., et al. 2007, \nat, 447, 183
\bibitem[Knutson et al.(2008)]{knutson08} Knutson, H. A., et al. 2008, \apj, 673, 526
\bibitem[Knutson et al.(2009a)]{knutson09a} Knutson, H. A., et al. 2009a, \apj, 690, 822 
\bibitem[Knutson et al.(2009b)]{knutson09} Knutson, H. A. et al. 2009b, \apj, 703, 769
\bibitem[Knutson et al.(2009c)]{knutson09c} Knutson, H. A., et al. 2009c, \apj, 703, 769
\bibitem[Knutson et al.(2010)]{knutson10} Knutson, H. A., Howard, A. W., \& Isaacson, H. 2010, \apj, 720, 1569
\bibitem[Kundurthy et al.(2011)]{kundurthy10} Kundurthy, P. et al. 2011, ApJ, 731, 123
\bibitem[Kurucz(1979)]{kurucz79} Kurucz, R. 1979, \apjs, 40, 1
\bibitem[Kurucz(1994)]{kurucz94} Kurucz, R. 1994, \emph{Solar Abundance Model Atmospheres for 0, 1, 2, 4, and 8 km/s} CD-ROM No. 19 (Smithsonian Astrophysical Observatory, Cambridge, MA, 1994)
\bibitem[Kurucz(1999)]{cdrom26} Kurucz, R. 1999, \emph{H$_2$O linelist from Partridge \& Schwenke (1997), part 2 of 2.} CD-ROM No. 26 (Smithsonian Astrophysical Observatory, Cambridge, MA, 1994)
\bibitem[Kurucz(2005)]{kurucz05} Kurucz, R. 2005, in Memorie Della Societa Astronomica Italiana Supplement, v. 8, p. 14
\bibitem[Langton \& Laughlin(2008)]{langton08} Langton, J., \& Laughlin, G. 2008, \apj, 674, 1106
\bibitem[Laughlin et al.(2009)]{laughlin09} Laughlin, G., et al. 2009, \nat, 457, 562
\bibitem[L\'eger et al.(2009)]{leger09} L\'eger, A., et a. 2009, \aap, 506, 287
\bibitem[Lewis et al.(2010)]{lewis10} Lewis, N. K., et al. 2010, \apj, 720, 344
\bibitem[Line et al.(2010)]{line10} Line, M. R., Liang, M. C., \& Yung, Y. L. 2010, \apj, 717, 496
\bibitem[Linsky et al.(2010)]{linsky10} Linsky, J. L., et al. 2010, \apj, 717, 1291
\bibitem[Loeb(2005)]{loeb05} Loeb, A. 2005, \apj, 623, L45
\bibitem[Lomb(1976)]{lomb76} Lomb, N. R. 1976, Ap\&SS, 39, 447
\bibitem[Madhusudhan \& Seager(2009)]{madhu09} Madhusudhan, N., \& Seager, S. 2009, \apj, 707, 24
\bibitem[Madhusudhan \& Seager(2011)]{madhu10} Madhusudhan, N., \& Seager, S. 2011, \apj, 729, 41
\bibitem[Madhusudhan \& Winn(2009)]{madhu09a} Madhusudhan, N., \& Winn, J. N. 2009, \apj, 693, 784
\bibitem[Mandel \& Agol(2002)]{mand02} Mandel, K. \& Agol, E. 2002, \apj, 580, L171
\bibitem[Maness et al.(2007)]{maness07} Maness, H. L., et al. 2007, \pasp, 119, 90
\bibitem[Markwardt(2009)]{markwardt09} Markward, C. B. 2009 \emph{in} D. A. Bohlender, D. Durand, \& P. Dowler, ed., 'Astronomical Society of the Pacific Conference Series' Vol. 411 of \emph{Astronomical Society of the Pacific Conference Series} pp. 251
\bibitem[Morales-Calderon et al.(2006)]{morales06} Morales-Calderon, M., et al. 2006, \apj, 653, 1454
\bibitem[Nettelmann et al.(2010)]{nettelmann10} Nettelmann, N., Kramm, U., Redmer, R., \& Neuh\"auser, R. 2010, \aap, 523, A26
\bibitem[O'Donovan et al.(2010)]{odonovan10} OÕDonovan, F. T., et al. 2010, \apj, 710, 1551
\bibitem[Orosz \& Hauschildt(2000)]{orosz00} Orosz, J. A., \& Hauschuldt, P. H. 2000, \aap, 364, 265
\bibitem[P\'al et al.(2010)]{pal10} P\'al, A., et al. 2010, \mnras, 401, 2665
\bibitem[Partridge \& Schwenke(1997)]{AmesH2O} Partridge, H., \& Schwenke, D.~W., 1997, \jcp, 106, 4618
\bibitem[Pont et al.(2008a)]{pont08a} Pont, F., et al. 2008a, \mnras, 385, 109
\bibitem[Pont et al.(2008b)]{pont08} Pont, F., et al. 2008b, \mnras, 393, L6
\bibitem[Queloz et al.(2009)]{queloz09} Queloz, D., et al. 2009, \aap, 506, 303
\bibitem[Rabus et al.(2009)]{rabus09} Rabus, M., et al. 2009, \aap, 494, 391
\bibitem[Rauscher et al.(2007)]{rauscher07} Rauscher, E., et al. 2007, \apj, 664, 1199
\bibitem[Reach et al.(2005)]{reach05} Reach, W. T. et al. 2005, PASP, 117, 978
\bibitem[Ribas et al.(2008)]{ribas08} Ribas, I., Font-Ribera, A., \& Beaulieu, J.-P. 2008, \apj, 677, L59
\bibitem[Rogers \& Seager(2010)]{rogers10} Rogers, L. A. \& Seager, S. 2010, \apj, 712, 974
\bibitem[Rothman et al.(2005)]{rothman05} Rothman, L. S., et al. 2005, J. Quant. Spec. \& Rad. Transfer, 96, 139
\bibitem[Scargle(1982)]{scargle82} Scargle, J. D. 1982, \apj, 263, 835
\bibitem[Shabram et al.(2011)]{shabram10} Shabram, M., Fortney, J. J., Green, T. P., \& Freedman, R. S. 2011, \apj, 727, 65
\bibitem[Sharp \& Burrows(2007)]{sharp07} Sharp, C. M. \& Burrows, A. 2007, \apjs, 168, 140
\bibitem[Shporer et al.(2009)]{shporer09} Shporer, A., et al. 2009, \apj, 694, 1559
\bibitem[Spiegel et al.(2010)]{spiegel10} Spiegel, D. S., Burrows, A., Ibgui, L., Hubeny, I., \& Milsom, J. A. 2010, \apj, 709, 149
\bibitem[Sterne(1940)]{sterne40} Sterne, T. E. 1940, Proc. National Academy of Sciences, 26, 36
\bibitem[Stevenson et al.(2010)]{stevenson10} Stevenson, K. B., et al. 2010, \nat, 464, 1161
\bibitem[Swain et al.(2008)]{swain08} Swain, M. R., Vasisht, G., \& Tinetti, G. 2008, \nat, 452, 329
\bibitem[Todorov et al.(2010)]{todorov10} Todorov, K., et al. 2010, \apj, 708, 498
\bibitem[Torres(2007)]{torres07} Torres, G. 2007, \apj, 671, L65
\bibitem[Vidal-Madjar et al.(2003)]{vidal03} Vidal-Madjar, A., et al. 2003, \nat, 422, 143
\bibitem[Werner et al.(2004)]{wern04} Werner, M.~W. et al., 2004, \apjs, 154, 1
\bibitem[Williams et al.(2006)]{williams06} Williams, P. K., Charbonneau, D., Cooper, C. S., Showman, A. P., \& Fortney, J. J. 2006, \apj, 649, 1020
\bibitem[Winn et al.(2007)]{winn07} Winn, J. N., et al. 2007a, \apj, 133, 1828
\bibitem[Winn et al.(2007b)]{winn07b} Winn, J. N., et al. 2007b, \aj, 134, 1707
\bibitem[Winn et al.(2008)]{winn08} Winn, J. N., et al. 2008, \apj, 683, 1076
\bibitem[Winn et al.(2010a)]{winn10a} Winn, J. N., Fabrycky, D., Albrecht, S., \& Johnson, J. A. 2010a, \apj, 718, L145
\bibitem[Winn et al.(2010b)]{winn10b} Winn, J. N., et al. 2010b, \apj, 723, L223
\bibitem[Zahnle et al.(2009)]{zahnle09} Zahnle, K., Marley, M. S., Freedman, R. S., Lodder, K., \& Fortney, J. J. 2009, \apj, 701, L20
\bibitem[Zahnle et al.(2010)]{zahnle10} Zahnle, K., Marley, M. S., \& Fortney, J. J. 2010, \apj, submitted, arXiv:0911.0728

\end{thebibliography}
\end{document}